\documentclass{aa}  
\usepackage{graphicx}
\usepackage{txfonts}
\usepackage{xcolor}
\usepackage{blindtext}
\usepackage{hyperref}

\begin{document} 

\title{Millimetre continuum from LBV stars and their environs\thanks{This work is based on observations carried out under project number 044-17 with the IRAM 30m telescope. IRAM is supported by INSU/CNRS (France), MPG (Germany) and IGN (Spain).}}

\subtitle{NIKA2 observations and Virtual Observatory data}

\author{J. R. Rizzo
          \inst{1}
          \and
          C. Bordiu\inst{2}
          \and
          A. Ritacco\inst{3}
          }

\institute{
              {ISDEFE, Beatriz de Bobadilla 3, 28040 Madrid, Spain.}
              \email{jrrizzo@isdefe.es}
         \and
            {INAF--Osservatorio Astrofisico di Catania, Via Santa Sofia 78, 95123 Catania, Italy}
         \and {Université Grenoble Alpes, CNRS, LPSC-IN2P3, 53, avenue des Martyrs, 38000 Grenoble,
France}
}

\date{Received May 10, 2025; accepted August 26, 2025}

\abstract
   {Luminous Blue Variables (LBVs) represent a brief transitional phase in the evolution of massive stars. Multi-wavelength studies of their circumstellar environments are essential to quantify their feedback at Galactic scales. Dominant emission mechanisms at millimetre wavelengths are, however, still poorly understood.}
   {Stellar winds, circumstellar dust, and ionised gas have not been explored together in the case of LBVs. We aim to study the millimetre continuum emission of Galactic LBVs to disclose the presence of these components, to describe their morphology and to measure their relevance in the mass and energy injection to the interstellar medium.}
   {We have used the NIKA2 continuum camera at the IRAM 30-m radio telescope to observe and analyse 1.15 and 2~mm continuum from the LBVs \object{HD168607}, \object{HD168625}, \object{[GKF2010]MN87}, \object{[GKF2010]MN101}, and \object{G79.29+0.46}. We used Virtual Observatory to complement our observations with archival data from optical, infrared, mm- and cm-wavelengths. With this information, we have built complete spectral energy distributions (SEDs) for the five sources, covering six decades of the electromagnetic spectrum.}
   {All targets except MN87 were detected at both wavelengths, with features including compact sources, extended nebular emission, shells, and unrelated background structures. Spectral indices of compact sources are consistent with thermal emission from stellar winds. We modelled the SEDs and successfully reproduced the emission from stellar photospheres, circumstellar dust, thermal stellar winds and enshrouding {\sc Hii} regions. Our models, in agreement with previous literature results, reveal the presence of unresolved hot dust very close to the stars and provide the first estimates for the fundamental parameters of MN101.}
   {This pilot study highlights the great potential of millimetre continuum studies of LBVs and possibly other evolved massive stars. The mm spectral window bridges the far-IR and radio regimes and can disclose the relative contribution of dust and free-free emission in this kind of sources.}

\keywords{Radio continuum: stars --
                Stars: massive --
                Stars: evolution --
                Circumstellar matter --
                Virtual Observatory tools --
                Astronomical data bases
               }

\titlerunning{Millimetre continuum from LBVs and their environs}
\authorrunning{Rizzo et al.}

\maketitle

%

\section{Introduction}
\label{sec:introduction}

The late evolutionary stages of stars of high mass have a notorious influence in the shaping and evolution of the Milky Way. As these massive stars evolve through a series of transitional phases to their endpoint as core-collapse supernovae \citep{lan94}, their intense UV fields and conspicuous mass-loss deeply transform the dynamics, structure, and chemical composition of their environs. Among these stages, a short interlude of \mbox{$\sim10^4$\,yr} known as the Luminous Blue Variable phase (hereafter LBV) stands out as a major contributor in terms of its radiative and mechanical output \citep{hum94}, thanks to high mass-loss rates ($\dot M$ up to \mbox{$10^{-4}$ M$_\odot$ yr$^{-1}$}) and occasional eruptions (shedding out up to some $\sim$10 M$_\odot$ in extremely short timescales). These episodes of enhanced mass-loss, that may rival the energetic output of supernovae \citep{smi13}, leave an indelible footprint in the outskirts of these stars, giving rise to complex and heterogeneous circumstellar nebulae of dust and gas \citep{wei01,wei11}.

This circumstellar material has been probed by diverse observational strategies. Infrared continuum and spectroscopy revealed the existence of different circumstellar dust populations around some stars, traced by successive mass-loss events \citep{jim10,vam13,vam15,bue17} and yielded valuable insights on the chemistry and dust lifecycle around these sources \citep{gul20}. Continuum emission at centimetre wavelengths traced the distribution of ionised gas around the stars and detected their thermal stellar winds, which enabled accurate estimates of their mass-loss budget \citep{dun02,uma10,ing16} and, under some assumptions, their present-day mass-loss rate \citep{uma05,uma12}. Molecular spectroscopy studies of the mm window revealed molecular structures tracing swept-up material and past events of mass ejection. This component is critical to our understanding of the physical processes that drive the evolution of the circumstellar material (i.e., energy and mass-loss balance) \citep{riz08,bor19, bor21} and its chemical complexity \citep{riz14,mor20,bor22,riz23}. 

The mm region of the electromagnetic spectrum is, a priori, the one where cold dust and ionised gas can contribute significantly through continuum emission. Observations in this range are therefore needed to characterize and constrain the relative importance of each component. Despite its potential relevance, there has been limited research on the (sub) millimetre continuum emission of LBV stars. In the Galaxy, we only found observational studies in \object{G79.29+0.46} \citep{hig94}, \object{Wd1-243} \citep{fen18}, \object{$\eta$~Car} \citep{bor19}, and \object{AG~Car} \citep{bor21}. In the first two cases, there are only flux measurements at the central source, while in AG~Car, it was detected the central source and weak emission from the surrounding nebula. In $\eta$~Car, the mm continuum maps revealed both the stellar wind and the extended ejecta from the 1890s outburst.

A handful of extragalactic LBV stars in the Large Magellanic Cloud have also been observed: S61, RMC 127, and RMC 143 \citep[][respectively]{agl17a, agl17b, agl19}, all of them known to harbour dense circumstellar nebulae visible at infrared or centimetric wavelengths. In the case of S61, neither the central star nor the shell were detected; RMC 127 appeared as a point-like source but the nebula was not detected; and RMC 143 was clearly detected, surrounded by a massive resolved nebula of dust and ionised gas, different in morphology from the circumstellar material observed at longer wavelengths. 

The disparity of results in such a small source sample underlines the relevance of the (sub) millimetre characterization as a valuable tool in the study of these objects. Indeed, the millimetre regime is the range where thermal dust and bremsstrahlung emission compete, so observations in this band are fundamental to disentangle their contribution to the overall mass-loss budget.

To mitigate this lack of sampling and gain a better understanding of the processes governing emission in this part of the spectrum, we used the NIKA2 continuum camera at the IRAM 30-m telescope and carried out a pilot study of the 1 and 2~mm continuum emission of five Galactic LBV stars (see Table~\ref{tab:sources}): 

\begin{itemize}
    \item \object{HD168607}. Located at a distance of $\sim$1.8 kpc \citep{gai22}, this source is a confirmed LBV star of spectral type B9Ia+ \citep{cla12} that shows no signs of circumstellar material \citep{hut94}.
    
    \item \object{HD168625}. Projected in the sky at $\sim 1$\,arcmin, and slightly closer to \mbox{HD168607}, this star is a supergiant of spectral type B8Ia+ \citep{cla12}. It is enshrouded in an hourglass-shaped nebula likely produced as a result of binary interaction \citep{mar16,mah16}. It is also recognized as a near-perfect analogue of the circumstellar environment of SN1987A \citep{smi07}. Since substantial variability is not fully demonstrated \citep{cla12}, it remains as a candidate LBV, being one of the less luminous stars of its kind \citep{smi04}.
    
    \item \object{$[$GKF2010]MN87} (hereafter MN87). This source was first discovered through its associated bright and asymmetric shell of radius \mbox{28\,arcsec} in the MIPSGAL survey \citep{miz10}. It was promptly identified as as a massive star based on infrared spectroscopy \citep{wac10,gva10}. Later it was proposed as a LBV candidate on the basis of high resolution radio observations \citep{ing14,ing16}. The nebula central source is compatible with a thermal stellar wind and a clumpy ionised shell. Its distance is unknown. Assuming a radius of \mbox{0.3 -- 0.6\,pc}, typical of other LBV shells, the most likely range of distances is 2 -- 4\,kpc.
    
    \item \object{$[$GKF2010]MN101} (hereafter MN101). This source was also discovered in MIPSGAL as a $\sim$20 arcsec diameter nebula and proposed as a LBV candidate by NIR spectroscopy of the central source \citep{wac11,fla14}. \citet{ing14,ing16} studied its radio emission and revealed a central source with a canonical thermal wind, and an ionised nebula co-spatial with the infrared, characterised by a negative spectral index. Finally, \cite{bor19} detected a CO torus wrapping the infrared nebula, with a low [$^{12}$CO/$^{13}$CO] ratio indicative of CNO-processed material. The distance to this source is uncertain, with the most likely range of geometric distances estimated from Gaia parallaxes \citep{bai21} being 3.0--6.2 kpc.
    
    \item \object{G79.29+0.46}. This source is an archetypal LBV candidate \citep{wat96} surrounded by a series of nested infrared shells  \citep{voo00, jim10} associated with successive mass-loss episodes. The innermost shell has a ionised counterpart detected in radio continuum observations \citep{uma11,agl14}, and a molecular counterpart traced by CO emission \citep{riz08}, which partially interacts with the nearby IRDC G79.3+0.3 \citep{pal14}. Some warm hotspots of ammonia have also been reported in and within the shell \citep{riz14}. The distance to G79.29+0.46 is \mbox{1.75\,kpc} \citep{gai22}.
\end{itemize}

The continuum maps delivered by the NIKA2 camera have fields of view (FOV) of several arcmin, enabling the simultaneous study of the star and their surroundings, with the potential to reveal the presence of multiple co-existing emission mechanisms. In Sect.~2, we describe the observations. In Sect.~3, we present the main results, including the overall maps, zoom-ins around the stars and a study of the millimetre spectral indices. In Sect.~4, we complement the information by the construction of spectral energy distribution (SED) diagrams from the optical to radio wavelengths. In Sect.~5, we identify and model the components present in the SEDs, which allowed us to draw some general ideas about the dominant emission mechanisms. We conclude with a general discussion about the survey and the individual sources in Sect.~6.

\begin{table*}
\caption[]{Sources observed.}
\label{tab:sources}
\centering
\begin{tabular}{lcrrrlcc}
\hline\hline
\noalign{\smallskip}
\multicolumn{1}{c}{Name} & 2MASS name & \multicolumn{1}{c}{RA(J2000)} & \multicolumn{1}{c}{Dec(J2000)} & \multicolumn{1}{c}{distance} & \multicolumn{1}{c}{Sp.~type} & Ref. & \multicolumn{1}{c}{CSM size} \\
&& \multicolumn{1}{c}{deg} & \multicolumn{1}{c}{deg} & \multicolumn{1}{c}{kpc} &&& \multicolumn{1}{c}{arcsec} \\
\noalign{\smallskip}
\hline
\noalign{\smallskip}
HD168607         & J18211489-1622318 & 275.3120 & $-$16.3755 & $1.84\pm0.08$   & B9Ia+  & 1 & --- \\
HD168625         & J18211955-1622260 & 275.3315 & $-$16.3739 & $1.53\pm0.06$   & B8Ia+  & 1 & 40  \\
MN87             & J18422247-0504300 & 280.5936 &  $-$5.0750 & (2.0--4.0)      & ---    &   & 28  \\
MN101            & J19062457+0822015 & 286.6024 &  $+$8.3671 & (3.0--6.2)      & ---    &   & 20  \\
G79.29+0.46      & J20314228+4021591 & 307.9262 & $+$40.3664 & $1.75\pm0.18$   & B:I[e] & 2 & 160 \\
\noalign{\smallskip}
\hline
\end{tabular}
\tablefoot{
Distances derived from Gaia DR3 \citep{gai22} in HD168607, HD168625, and G79.29+0.46. For MN87 the quoted distance range is derived from the angular size of the shell, assuming a physical size of 0.3--0.6 pc, as observed in other LBV stars \citep{wei11}. For MN101, the quoted distance range is the most likely range of geometric distances derived from Gaia parallaxes \citep{bai21}. Quoted references are the papers used for the spectral classification of the target stars. CSM sizes are the angular diameters obtained by visual inspection of the images depicted in the papers cited in Sect.~\ref{sec:introduction}.
\tablebib{(1)~\citet{cla12}; (2)~\citet{voo00}}
}
\end{table*}

\section{Observations}
Observations were done with the NIKA2 continuum camera \citep{per20} available at the IRAM 30m radio telescope, during three consecutive runs in October 2017 (project 044-17; P.I.~R.~Rizzo). Weather was stable during all the sessions, with measured opacities at \mbox{225\,GHz} between 0.2 and 0.3, equivalent to precipitable water vapour between 3 and \mbox{5\,mm}.

For the four regions of interest we have performed a total of 126 individual on-the-fly maps with duration of three to five minutes each. Scanning angles were 0, 45 and 90 degrees to avoid stripping effects in the reconstruction of the final maps. Additional 15\% of extra size has been added to the definition of the maps to account for a better reconstruction of the background noise. Pointing was regularly checked, with corrections always below 4~arcsec. The FOV for all sources but G79.29+0.46 is \mbox{6.5\,arcmin}. In G79.29+0.46, we observed with 8.6 arcmin of FOV in order to include the previously known infrared shells and the nearby IRDC.

The NIKA2 camera has three arrays (two at 260\,GHz and one at 150\,GHz) filled with thousands of Kinetic Inductance Detectors (KIDs), providing an accurate combination of the responses from off-source detectors to reconstruct the noise template to be subtracted from the data.

To produce the final images, we have used a data analysis pipeline based on MOPSIC \citep{zyl13} and developed by the NIKA2 collaboration prior to the first public release of PIIC\footnote{\url{https://www.iram.fr/~gildas/dist/piic.pdf}}. In order to preserve large scale features, we have used an iterative procedure. In the first iteration, the software accounts for the absolute calibration of the flux, the opacity correction and merge together all the maps collected. As output of this first iteration we have a first attempt of the total map. This is used to estimate the S/N on the map and extrapolate the region with maximum brightness. The extrapolated structure of the source is used as a mask that is introduced in a second round of iteration. At the end of this latter we obtain a ``convergence'' map, which results from the difference between the two previous ones. If the convergence map has a brightness greater than a well defined threshold in the central region, the process continues and a third iteration starts. This method is based on minimising the residuals on the final map and allows to recover the largest scales in case of extended emission.

\begin{figure}
    \centering
    \includegraphics[width=1\linewidth]{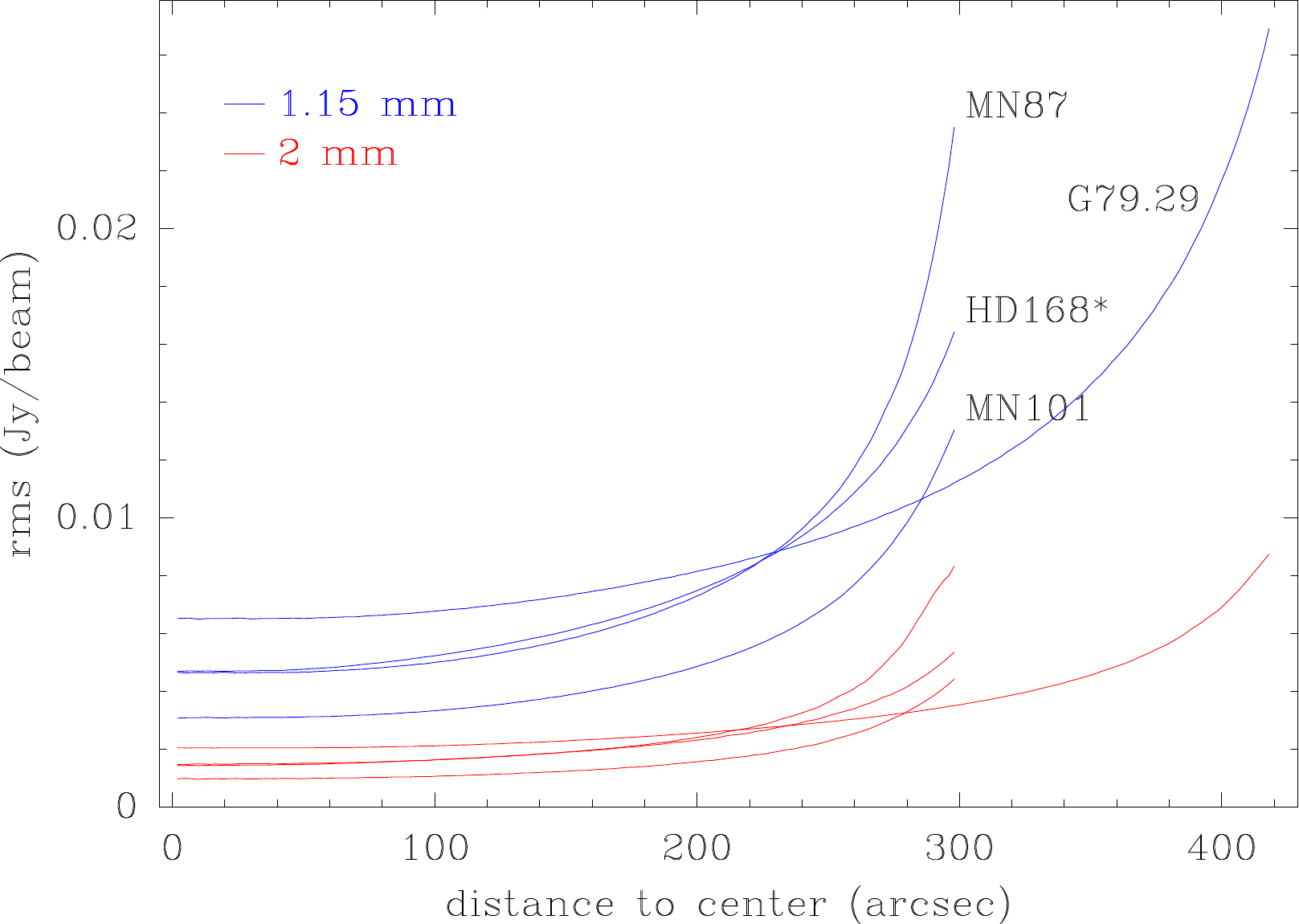}
    \caption{Noise map of the fields observed. The $rms$ noise was computed in concentric rings of \mbox{2\,arcsec} wide. The noise in the 1.15~mm maps are plotted in blue, while the one corresponding to the 2~mm maps are in red. Field names (abridged) are indicated near the 1~mm curves.
    }
    \label{fig:noise}
\end{figure}

The $rms$ map values of the four fields are depicted in Fig.~\ref{fig:noise}. The values were computed as averages over concentric rings of 2~arcsec wide. At 1.15~mm (260\,GHz), the noise map is rather uniform up to \mbox{120 -- 140\,arcsec} of the centre, while at 2~mm (150\,GHz) it goes well beyond 200\,arcsec.

Finally, we smoothed the maps with a Hanning $3\times3$ kernel to improve the S/N. The resulting angular resolution (half power beam width, HPBW) after this step was 12.5 and 18.5 arcsec, at 1.15 and 2~mm, respectively. With the smoothing, the $rms$ has been reduced by a factor around 0.36 in all cases.

\begin{figure*}[ht]
\centering
\includegraphics[width=0.95\linewidth]{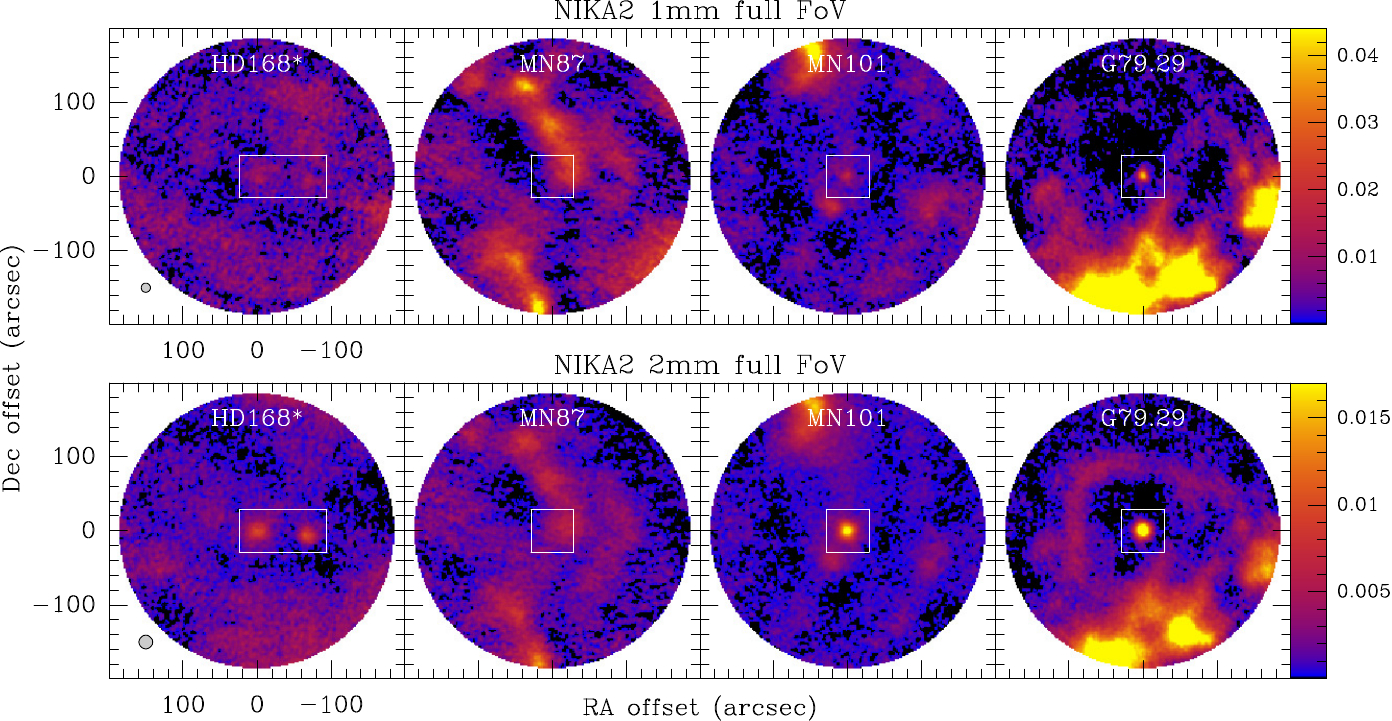}
\caption{Sky maps of the mm-continuum emission in the four fields observed with NIKA2. Maps at 1.15~mm are at the top row, while the 2~mm maps are on the bottom one. The field which contains HD168607 and HD168625, labelled HD168*, is centered in HD168625. Colour scales --indicated or the right of each row-- are in Jy~beam$^{-1}$. HPBWs are drawn near the bottom left corners of the HD168* images. Central areas indicate the zoomed region displayed in Fig.\,\ref{fig:zoom}.} 
\label{fig:fields}
\end{figure*}

\section{NIKA2 results}
\subsection{Emission from the stars and their surroundings}
\label{sec:emission}

\begin{figure*}[ht]
\centering
\includegraphics[width=0.95\linewidth]{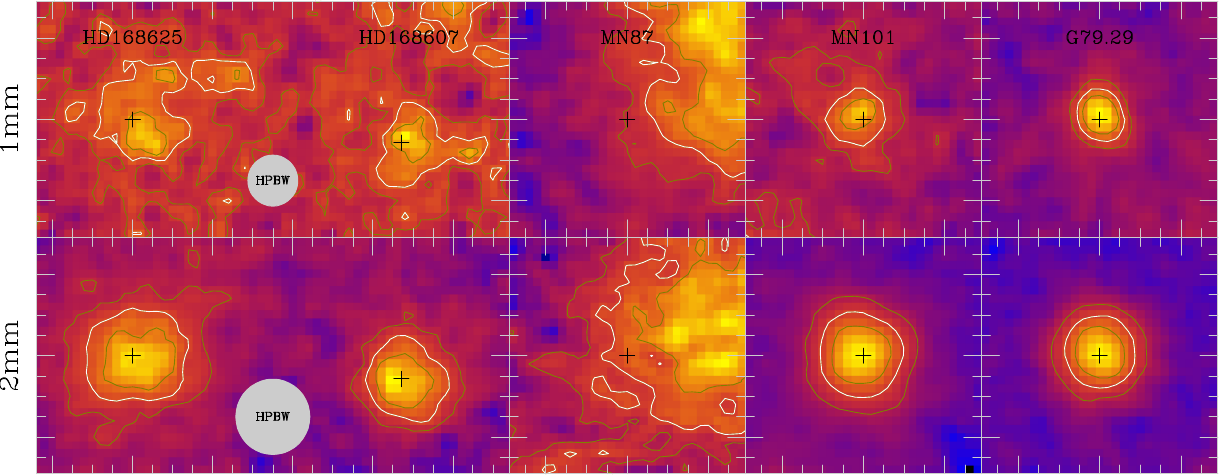}
\caption{Zoomed sky maps around the central sources. The regions correspond to those indicated in Fig.~\ref{fig:fields}. To span the full dynamic range, colour scale is different for each of the maps. Contours correspond to 30, 50 (highlighted in white), and 70~\% of the maximum value. HPBWs are indicated near the bottom of the HD168* images. The contours at 50~\% are highlighted to facilitate a visual comparison with the beam, and give an idea about the point-like nature of the central sources. } 
\label{fig:zoom}
\end{figure*}

The final processed maps are presented in Fig. \ref{fig:fields}. Overall, we notice emission from the stars, circumstellar structures and other likely unrelated features. A close-up of the emission at the star positions is depicted in Fig.~\ref{fig:zoom}, with the zoom at 1.15~mm and 2~mm depicted at the top and bottom rows, respectively. To facilitate a visual inspection of the point-like nature of the sources, we highlighted a contour corresponding to 50~\% of the continuum peak, together with the HPBW at each frequency. 

HD168625 and HD168607 are detected in both bands, but more clearly at 2 mm. The entire field is relatively devoid of significant emission except around the stars. In the case of HD168625 the source appears slightly resolved. At 1.15~mm, it shows a faint extended plateau of emission, peaking towards the south-west and possibly linked to the complex, hourglass-shaped nebula that enshrouds the star at infrared wavelengths \citep{smi07}. At 2~mm, however, the emission is rather circular, with the maximum in good positional agreement with the star. On the other hand, HD168607, which lacks reported circumstellar material, is  unresolved at 2~mm, and very probably also at 1.15~mm, although the low S/N prevents a more in-depth analysis.

MN87 is not detected in neither band. The field is dominated by a series of filaments, brighter at 1.15\,mm. These structures are well correlated with the dust bands seen in images at 250 and 500~$\mu$m from Herschel/SPIRE \citep{mol10}.

The field of MN101 is dominated by a bright emission region towards the north, likely corresponding to a molecular cloud unrelated to the star. Towards the centre of the map, two components can be barely resolved: a central compact emission that matches the position and size of the infrared nebula, possibly embedded in a plateau, and a more diffuse component located about 30~arcsec to the south-east, corresponding to the brightest region of the CO structure reported by \cite{bor19}. While the two components have a similar brightness at 1.15~mm, the central source dominates at 2~mm.

In the field of G79.29+0.46, the bright IRDC stands out, extending from the south-east to the west. The central source is well detected in the two bands, with notably brighter emission at 2~mm. A shell-like structure surrounding the star from the north-east is also detected, faintly at 1~mm and clearly at 2~mm. This structure has an approximate radius of 100 arcsec, meaning it is most likely associated with the innermost infrared dust shell, as visible in \textit{Spitzer} images \citep{jim10}. This millimetre shell is bounded by the CO shells reported by \cite{riz08}.

The Table \ref{tab:fitting} lists the flux densities of the central sources at 1.15 and 2~mm along with their associated uncertainties, centroids and sizes, all of them derived from Gaussian fitting. The last column also shows the spectral indices, which are discussed in the next Section.

\begin{table*}
\caption{Two-dimensional Gaussian fitting.}
\label{tab:fitting}
\centering
\begin{tabular}{lcc@{\extracolsep{3mm}}ccc@{\extracolsep{3mm}}ccc@{\extracolsep{3mm}}c}
\hline\hline
\noalign{\smallskip}
Source             & RA$_\mathrm{peak}$ & Dec$_\mathrm{peak}$ & \multicolumn{3}{c}{$\lambda=1.15~\mu$m} & \multicolumn{3}{c}{$\lambda=2~\mu$m} & $\alpha_{\rm mm}$ \\
\noalign{\smallskip}
\cline{4-6} \cline{7-9}
\noalign{\smallskip}
&&& $r_\mathrm{maj}$ & $r_\mathrm{min}$ & $S_\nu$ &  $r_\mathrm{maj}$ & $r_\mathrm{min}$ & $S_\nu$ \\
& deg & deg & arcsec & arcsec & mJy & arcsec & arcsec & mJy \\
\noalign{\smallskip}
\hline
\noalign{\smallskip}
HD168607           & 275.3240 & -16.3756 & 14.1$\pm$0.6  & 14.2$\pm$5.8  & 13.2$\pm$8.6  & 19.8$\pm$1.8 & 18.6$\pm1.6$ & 10.1$\pm$1.5 & 0.49\\
HD168625           & 275.3308 & -16.3737 & 21.1$\pm$10.1 & 21.0$\pm$10.0 & 25.1$\pm$15.6 & 28.1$\pm$3.9 & 24.5$\pm3.1$ & 16.6$\pm$3.1 & 0.75\\
MN87               & ---      & ---      & ---           & ---           & $<15$\tablefootmark{a}        & ---          & ---          & $<5$\tablefootmark{a} & --- \\
MN101              & 286.6027 &   8.3674 & 13.4$\pm$3.1  & 13.9$\pm$3.3  & 16.5$\pm$6.4  & 20.3$\pm$2.4 & 19.3$\pm2.2$ & 19.3$\pm$3.7 & -0.29\\
G79.29+0.46        & 307.9261 & -40.3667 & 13.6$\pm$0.6  & 11.9$\pm$0.5  & 42.0$\pm$3.2  & 18.3$\pm$0.4 & 18.1$\pm0.4$ & 25.0$\pm$1.0 & 0.94\\
\noalign{\smallskip}
\hline
\end{tabular}
\tablefoottext{a}{
Upper limits quoted for MN87 are three times the corresponding map noises.
}
\end{table*}

\subsection{Millimetre spectral indices}

We have computed the spectral indices of the central sources between the two NIKA2 frequencies ($\alpha_{\rm mm}$ hereafter) and following the convention $S_\nu\propto\nu^\alpha$. Despite the low S/N at \mbox{1\,mm}, it is overall evident that all the detected sources except MN101 exhibit values close to a thermal stellar wind ($\alpha_{\rm mm}\sim 0.6$), which is indeed expected for the strong winds driven by LBV stars.

\begin{figure*}[ht]
\centering
\includegraphics[width=0.95\textwidth]{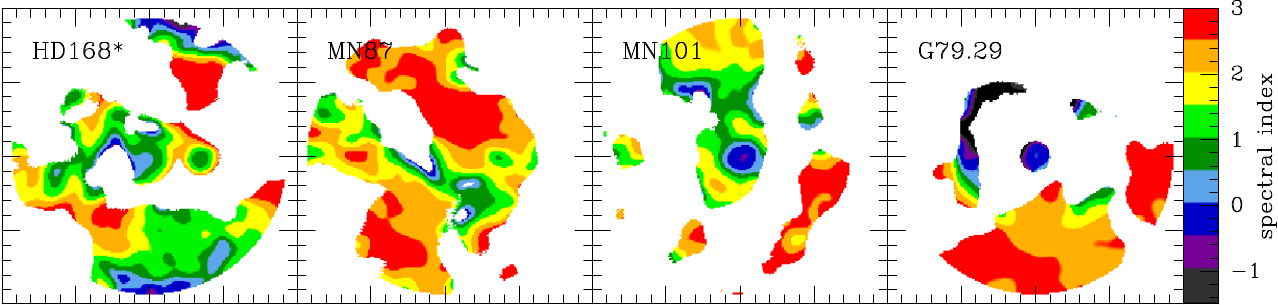}
\caption{
Spectral index maps between the two NIKA2 frequencies. The plotted field and marks are the same as in Fig.~\ref{fig:fields}. For each map, all pixels with brightness below 2$\sigma$ have been masked.
}
\label{fig:spindex}
\end{figure*}

On the other hand, we have built the maps of $\alpha_{\rm mm}$ for all the observed fields after convolving them to a common beam of \mbox{20\,arcsec}. We used only those pixels with brightness above a 2$\sigma$ threshold in the convolved maps. The resulting maps are shown in Fig.~\ref{fig:spindex}. They reveal a diverse range of features, likely due to multiple emission mechanisms at work.

\object{HD168625} shows a somewhat negative index, consistent with an ionised H\textsc{ii} region. As a consequence of the convolution, the emission at {1\,mm} is severely diluted and displays at the central position the contribution of the central source plus that of an additional component, probably close-in, young, stellar ejecta. On the contrary, \object{HD168607} displays slightly positive values of $\alpha_{\rm mm}$, more compatible with a thermal stellar wind, in line with the values reported in Table \ref{tab:fitting}.

In the case of MN87, the dominant emission arises primarily from filaments which cross the whole field. These features have high values of $\alpha_{\rm mm}$, typical of thermal dust emission. Considering also their morphology, we think that the mm-emission in this field is very likely arising from foreground/background Galactic clouds.

MN101 presents a more complex scenario, where the central region depicts negative values of $\alpha_{\rm mm}$. Again, the NIKA2 beam plays here a critical role: the size of the infrared nebula is comparable to the beam, so the negative spectral index likely captures the combined contribution of the stellar wind and the nebular material, known to show non-thermal emission at the cm regime \citep{ing16}. The nebula is embedded in a free-free plateau, and surrounded by thermal dust (in yellow) in good correspondence with well known molecular structures \citep{bor19}. The highest spectral indices in this enshrouding structure are located south-west of the star, in coincidence with the brightest part of the CO cloud.

G79.29+0.46 is perhaps the most puzzling object. The central region exhibits a rather negative $\alpha_{\rm mm}$, in contrast with the positive value derived for the central source in Table \ref{tab:fitting} (i.e., not convolved). While the fitted value is in relatively good agreement with the $\alpha=0.6$ derived by \cite{hig94} at 0.8, 1 and {1.2\,mm}, the negative value observed in the map is a consequence of the severe dilution of the 1mm central point source (around 0.39).

G79.29+0.46 is known to host a set of concentric shells at angular distances of {$100-200$\,arcsec} from the exciting star \citep{riz08, jim10, agl14}. The inner shell is nicely seen in Fig.~\ref{fig:spindex}, showing a remarkable stratification at its eastern side around position \mbox{(100'', --20'')}, with three distinct components; moving outward, there is a transition from negative values, possibly indicating shocked material, to flat values corresponding to the ionised shell, and finally to positive values associated with the thermal dust surrounding the shell. A secondary, less pronounced stratification is evident in the south-west region, where the shell is interacting with a nearby IRDC \citep{riz08, pal14}, and where we measure the highest values of $\alpha_{\rm mm}$ (red colour in Fig.~\ref{fig:spindex}).

\section{SED analysis}
\subsection{Building}

Aiming to study the relative contribution of the different physical processes which dominate the emission of our targets, we constructed their SEDs along virtually the whole electromagnetic spectrum. As we are interested in the study of the circumstellar gas and dust, we focused the analysis in the infrared, mm, and radio domains.

For each target star, we proceeded in a number of sequential steps:

\begin{itemize}
\item We constructed a Python script to retrieve all the photometry data available in VizieR. The script was validated after comparing its outputs with the VizieR photometry tool\footnote{\url{http://vizier.cds.unistra.fr/vizier/sed/beta/}}. The script outputs have a  structure similar to that of VizieR, with the addition of the wavelength in microns.
\item We initially ran the script with a search radius of 3~arcsec. 
\item In order to check possible data lost beyond 3~arcsec (typically in infrared to cm wavelengths), we performed a second search by changing the search radius to 30~arcsec.
\item We removed duplicated information, giving more priority to original data and to data with flux uncertainties included, instead of compilation of other surveys and data without quoted uncertainties.
\item In the VizieR database\footnote{\url{https://vizier.cds.unistra.fr/viz-bin/VizieR}}, we checked all the literature information about the targets. This step was performed to incorporate possible data from catalogues not included in the photometry tool.
\item We used Simbad service\footnote{\url{http://simbad.cds.unistra.fr/simbad/}} to check and incorporate data from observational papers not included in VizieR database.
\end{itemize}

The Table~\ref{tab:catalogues} (in Appendix \ref{app:surveys}) lists all the datasets used to build the SEDs. Besides the catalogue names, the VizieR table number, the wavelength range and the bibliographic reference are also included. The Appendix~\ref{app:notes} enumerates a series of additional information about each of the stars, noticed in our tailored search for multi-wavelength photometry.

\subsection{Clarification on particular datasets}
\label{sec:clarification}

\paragraph{Gaia.} 
With the Gaia DR3 photometry data, we additionally checked the consistency of the G band flux with respect to the BP/RP ones. This effect is a consequence of different background estimates for G and for BP/RP bands\footnote{\url{https://gea.esac.esa.int/archive/documentation/GEDR3/Data\_processing/chap\_cu5pho/cu5pho\_sec\_photProc/cu5pho\_ssec\_photVal.html}}. 
While the G flux is determined by fitting a profile function, the BP and RP fluxes are determined by aperture photometry. The different approach may be critical in our case, where the targets are often very red due to surrounding warm dust and, at the same time, affected by contamination of other (bluer) stars due to their low Galactic latitude. The consistency is measured by the so-called excess factor, defined as the sum of fluxes in BP and RP bands, divided by the flux in G band; this factor is slightly above one in most cases. However, we noted very large excess factors in all the sources (1.41, 1.38, 2.14, 2.17, and 1.83 for the sources ordered as in Table \ref{tab:sources}). This fact might be interpreted by an incorrect measurement of the background in G (due to the inclusion of nebular material), rather than by contamination in BP and RP bands. In consequence, we did not consider the G band fluxes for the SED fitting.

\paragraph{GLIMPSE.} 
Its photometry is potentially affected by extended nebular emission, mainly due to the additional contribution of line emission and PAHs. A priori, this contamination may affect MN101, MN87 and HD168625, which are the sources where the nebular material is not completely detached from the star. Upon inspection of the fluxes, we confirm that MN87 and MN101 are not affected, as their nebular emission is only detected at 24$\mu$m \citep{miz10}. Contrarily, in HD168625,  the VizieR photometry tool only reports flux at 5.8\,$\mu$m, while the catalogue itself gives data at 3.6, 4.5, and \mbox{5.8\,$\mu$m} (all of them, though, well below the fluxes from other catalogues). Consequently, we removed these data points from the HD168625's SED.

\paragraph{Radio.} 
The radio data used to build the SEDs comes from both targeted high-resolution observations and all-sky surveys conducted at various radio facilities. Consequently, the angular resolution can vary by up to a factor of $\sim$10. While the highest-resolution data points measure directly the flux of the central stellar source, the lowest-resolution ones may include emission from both the star and the circumstellar material, except in cases where the star is completely detached from the nebula. In cases where we cannot separate the two components, the data points contain the sum of the flux densities corresponding to both the stellar wind and the surrounding nebula (modelled as bremsstrahlung, as explained in Sect. \ref{sec:fitting}).

\subsection{Resulting SEDs}

\begin{figure}[b]
\centering
\includegraphics[width=1\linewidth]{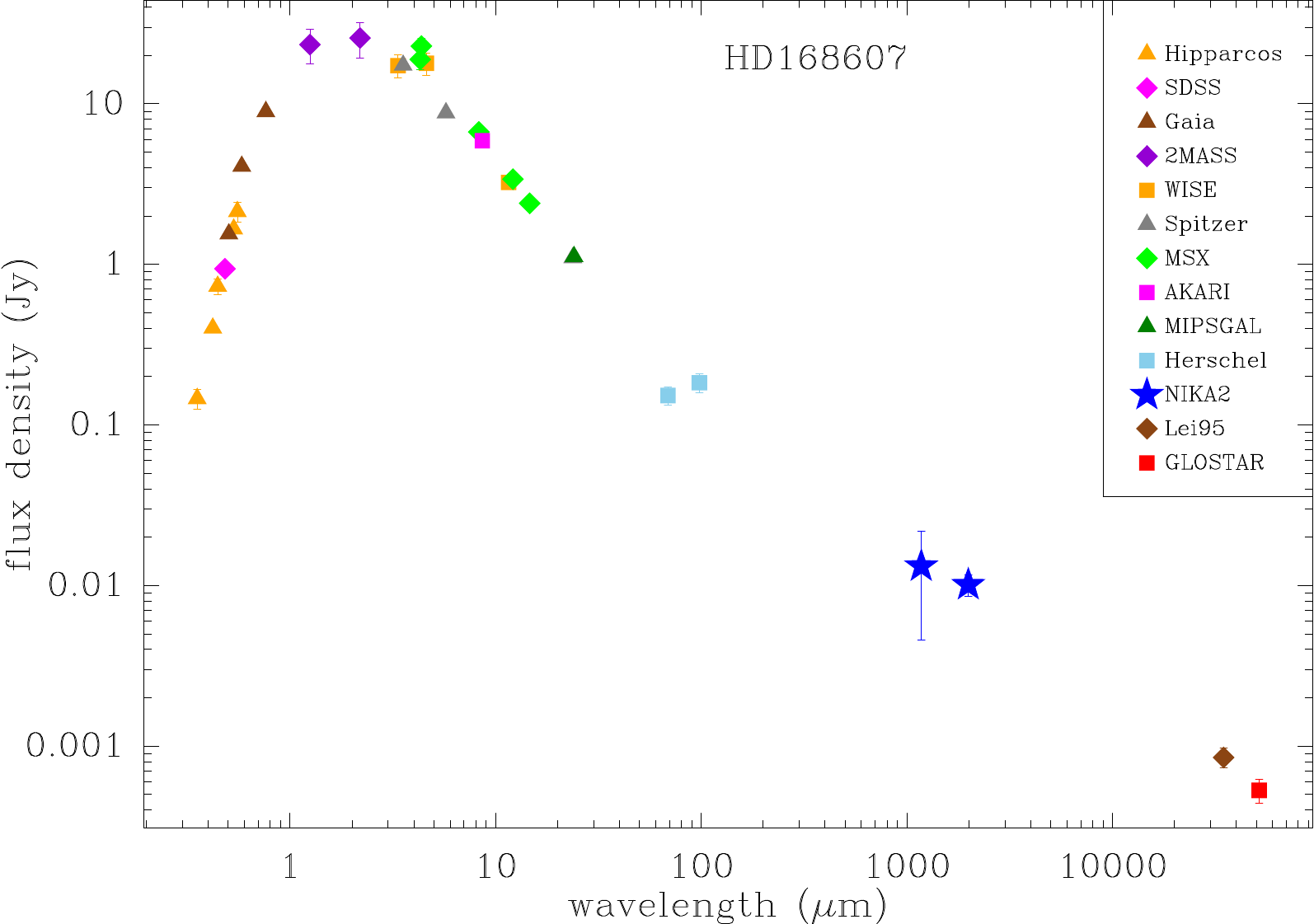}
\caption{Spectral energy distribution of HD168607, in logarithmic scale. Catalogues are indicated on top right. All points have their corresponding error bars plotted. The NIKA2 points are shown with blue stars.}
\label{fig:sed_HD168607}
\end{figure}

\begin{figure}[ht]
\centering
\includegraphics[width=1\linewidth]{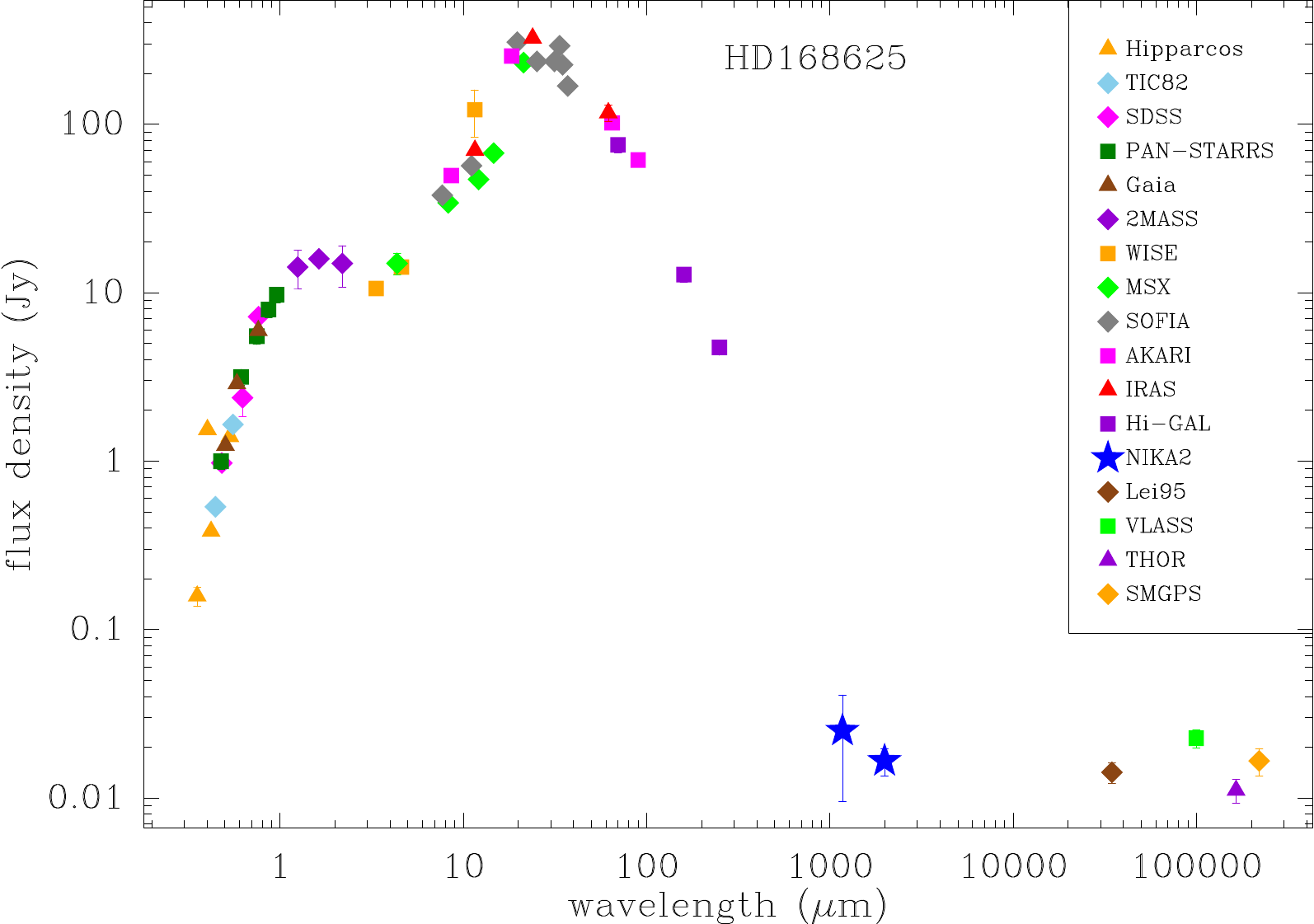}
\caption{Spectral energy distribution of HD168625.}
\label{fig:sed_HD168625}
\end{figure}

\begin{figure}[ht]
\centering
\includegraphics[width=1\linewidth]{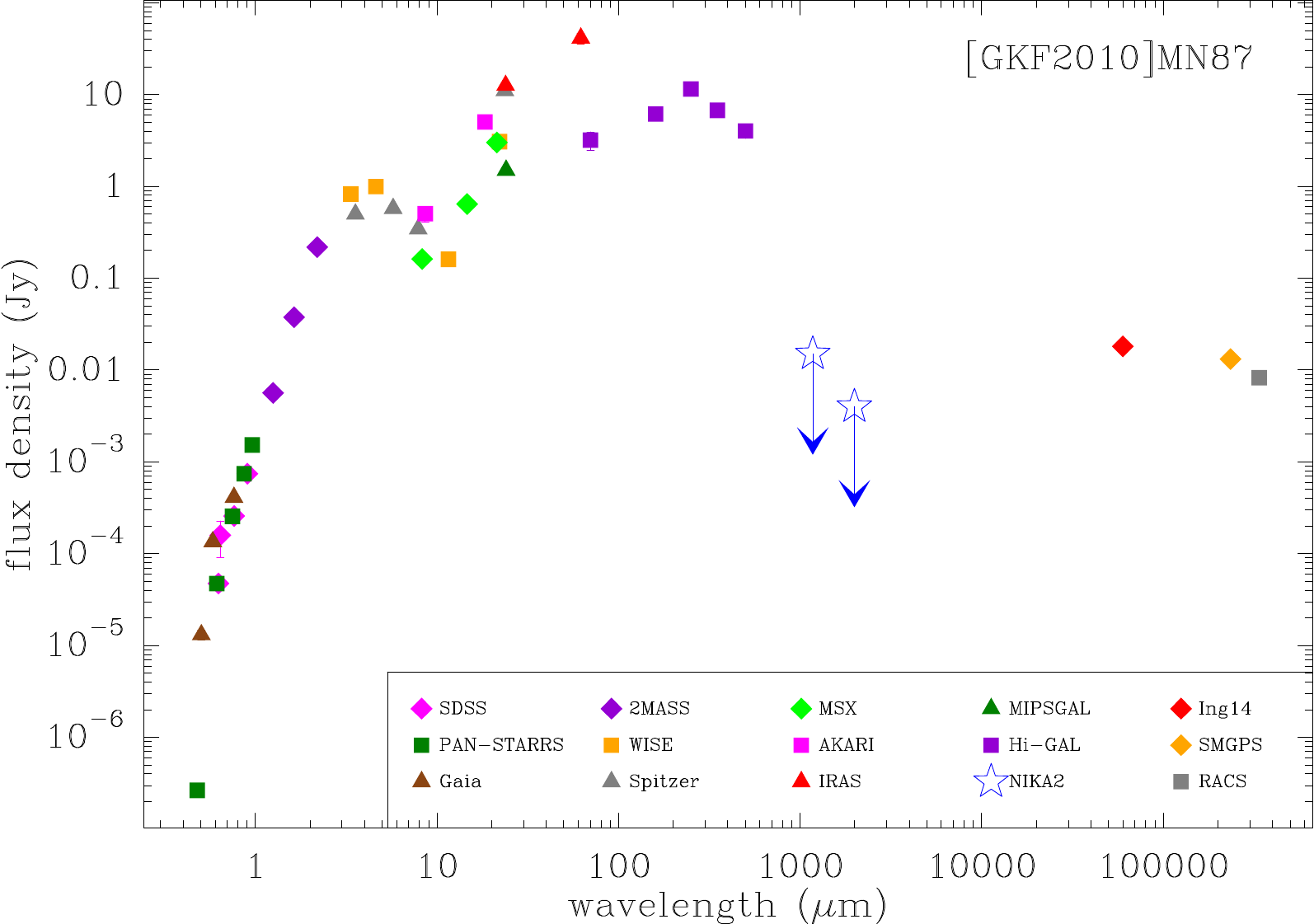}
\caption{Spectral energy distribution of [GKF2010]MN87.}
\label{fig:sed_MN87}
\end{figure}

\begin{figure}[ht]
\centering
\includegraphics[width=1\linewidth]{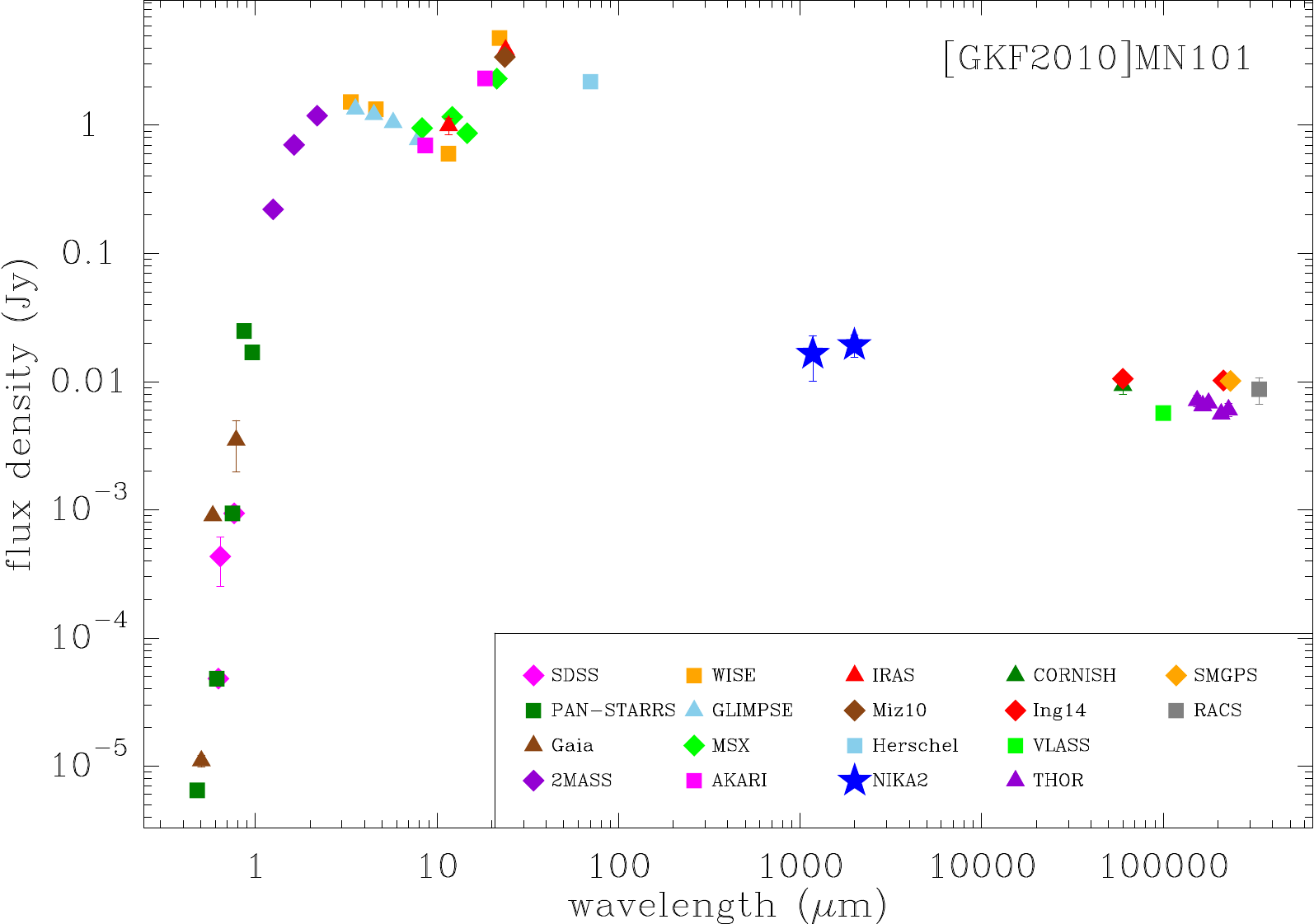}
\caption{Spectral energy distribution of [GKF2010]MN101.}
\label{fig:sed_MN101}
\end{figure}

\begin{figure}[ht]
\centering
\includegraphics[width=1\linewidth]{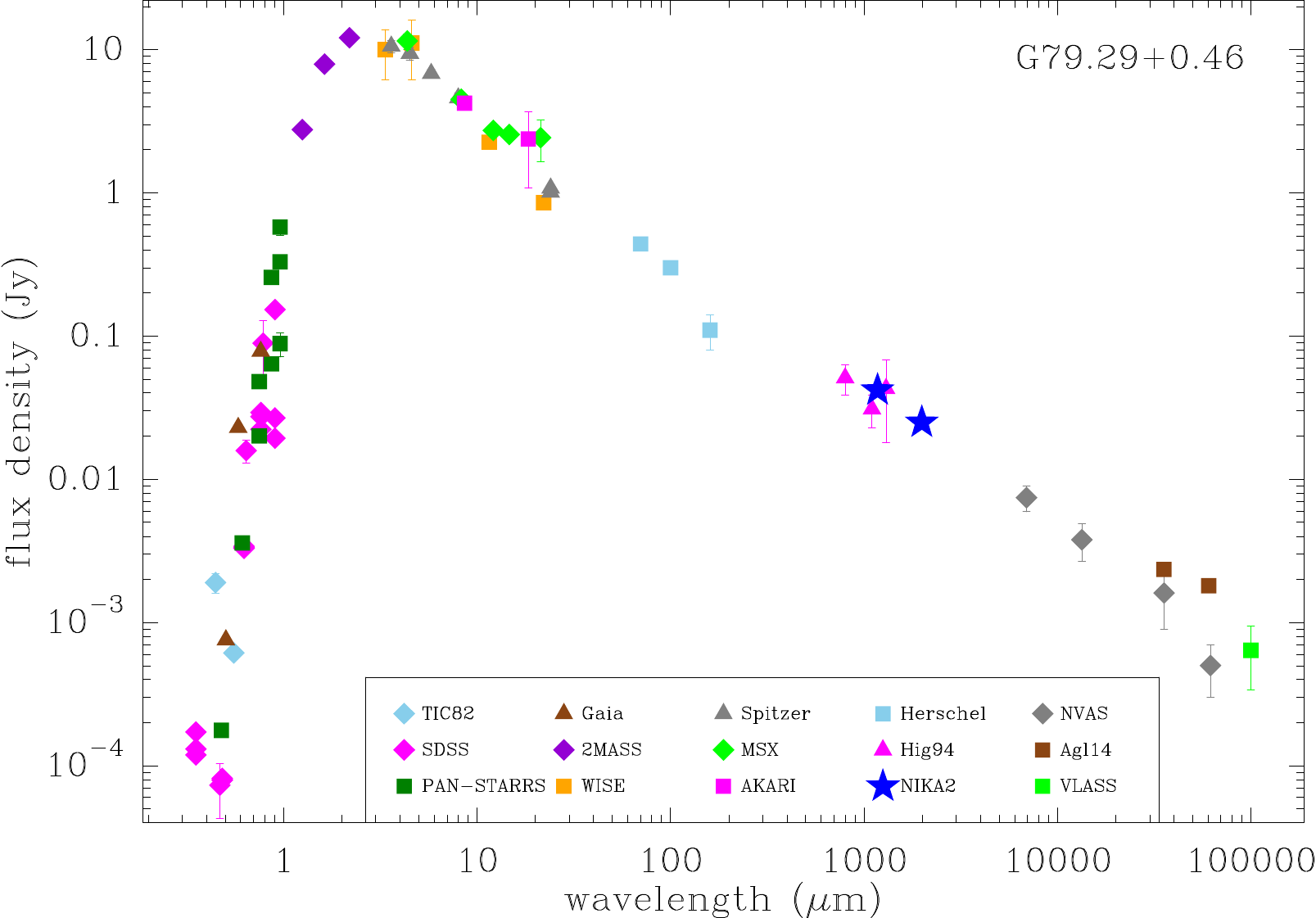}
\caption{Spectral energy distribution of G79.29+0.46.}
\label{fig:sed_G79}
\end{figure}

The resulting SEDs are depicted in Figs.~\ref{fig:sed_HD168607} to \ref{fig:sed_G79}. All of them are highly reddened, as their visual/near-IR peaks are in the range from 1 to $3\,\mu$m. This behaviour is indeed expected, considering the long distances of the stars, their location in the Galactic plane, and their nature as dust producers.

Roughly, we distinguish two types of SEDs. The first one is represented by \mbox{HD168607} (Fig. \ref{fig:sed_HD168607}) and G79.29+0.46 (Fig. \ref{fig:sed_G79}), where fluxes decrease monotonically from IR- to mm- to cm-wavelengths. In both cases, the NIKA2 fluxes are significantly higher than the radio ones (up to an order of magnitude). The two sources have no resolved ejecta structures falling within the NIKA2 beam (see Fig. \ref{fig:zoom}), so the measured flux is most likely the combined contribution of the stellar wind and close-in, unresolved dust.

The second type of SEDs is that of \mbox{HD168625} (Fig. \ref{fig:sed_HD168625}) and MN101 (Fig. \ref{fig:sed_MN101}). The two sources are immersed in conspicuous gaseous and dusty nebulae which contribute to the emission observed with NIKA2. Both SEDs are rather flat in the mm-cm range, with NIKA2 fluxes comparable to cm-wavelength ones (slightly below in HD168625 and slightly above in MN101), of the order of 10 mJy. In the infrared, both stars show a strong bump peaking at $\sim20-30\mu$m, probably indicating the presence of a large amount of warm circumstellar dust.

MN87 is a puzzling object, exhibiting strong infrared emission without a well-defined peak. This behaviour, not observed in the other four sources (for which we applied the same processing), is likely related to the different instrument apertures. In MN87, this effect seems a critical factor due to widespread Galactic contamination, as indicated by the dominance of SPIRE dust structures in the NIKA2 maps (see Sect. \ref{sec:emission}). At cm-wavelengths the hot star (more properly, its stellar wind) is clearly detected, with flux densities comparable to those of MN101 \citep{ing16}. 

\section{Multi-wavelength modelling}
\subsection{Fitting methodology}
\label{sec:fitting}

The environs of LBV stars are heavily affected by their strong UV radiation and powerful winds. Under adequate physical conditions, LBVs are also efficient dust producers. The continuum emission in these environments is dominated by a combination of thermal dust radiation and free-free emission produced by stellar winds and ionized nebular gas. Non-thermal emission may also be present under special circumstances, such as colliding winds in binary stars like \object{HD168625} \citep{mar16,mah16}, after mass ejection events, or in the presence of shocks.

The SEDs compiled and curated in the previous sections demonstrate that our targets depict all the above components with diverse significance at different wavelength ranges. Even more, the mm-range is key to constrain the contribution of the warm/cold dust and, at the same time, the highly ionized stellar wind. We therefore proceeded to model the observed SEDs aiming to evaluate the relative importance of these components. The models are intended to be as simple as possible, with the minimum number of parameters capable of providing robust estimates of the different physical components. Because the SEDs arise always from a single position and have not spectroscopic information, we do not attempt to provide fine details about the space distribution and velocity field of our targets. In some cases, we assumed reasonable physical quantities after checking their influence in the flux predictions. For each of the component modelled, Appendix \ref{app:models} thoroughly details these aspects.

The first step was to correct the stellar fluxes by interstellar extinction according to the Bayestar19 3D dust model \citep{gre19}. We assumed the distances quoted in Table \ref{tab:sources}, adopting 3 and 4~kpc for MN87 and MN101, respectively. We used the Python package {\tt dustmaps} \citep{gre18}, which provides extinction values for {\it grizy} and JHK bands, and interpolated the available photometry by a cubic spline. After this initial correction, we noted still significant remaining reddening in all the sources. This is indeed expected considering the presumed presence of circumstellar dust close to the extended star atmospheres. A follow-up determination of the total extinction was subsequently performed for wavelengths smaller than \mbox{4.6\,$\mu$m} \citep[for details see][]{gre19}, which allowed us to separate the interstellar and the circumstellar extinction at each optical/near-IR band. For this step, we considered effective temperatures ($T_\mathrm{eff}$) from 8000 to 20000 K, the range corresponding to all the known LBVs \citep{smi04}.

We modelled the extinction-corrected SEDs as a sum of independent flux contributors. We included four kinds of components: (1) a central star, modelled as a black body; (2) one or more circumstellar dust clouds, conceived as black or grey bodies at different temperatures; (3) stellar wind, modelled as a homogeneous and isotropic mass loss at a constant rate; and (4) free-free emission arising from a homogeneous ionized nebula surrounding the hot star. Appendix \ref{app:models} describes all these components in detail, including the free parameters and assumptions involved.

Modelling a SED from the optical to the radio bands (i.e., more than six decades) is challenging. The spectral regions have diverse coverage, aperture, and sampling; and all the flux measurements are also potentially affected by intrinsic variability of the sources, geometry, and composition, among other factors.

We followed an iterative fitting procedure, ensuring that each component is constrained progressively. First, we divided each SED into four spectral regions, bounded by those wavelengths where the contribution of two components are comparable. In general terms, the four regions are: (1) optical to a few microns, dominated by the contribution of the stellar photosphere; (2) near infrared up to some tens of microns, with contributions mainly from the star and a hot dust component; (3) mid-infrared up to the NIKA2 points, dominated by warm or cold dust components and free-free emission; and (4) mm- and cm-wavelengths, mostly dominated by ionised gas and/or the stellar wind.

We adopted a stepwise approach, starting from the optical region and progressively incorporating additional components:

\begin{itemize}
\item We started by fitting the optical part of the SED by the stellar photosphere alone. The parameter space representative of LBV stars ($T_{\rm eff}$ from \mbox{8\,000} to \mbox{20\,000~K} and stellar radius $R_{\rm star}$ from 30 to 300~R$_\odot$) was employed as a ``kick-off'' exploration, and later refined through $\chi^2$ minimisation.
\item Then, we proceeded to fit the near-infrared region. The star component was combined with an additional grey  body representing a hot dust component. The stellar parameters were kept within a narrow range around the previously determined, while the dust parameters were probed in ample ranges ($R_{\rm dust}$ from 0.5 to 20~au, ($T_{\rm dust}$ from 500 to \mbox{3\,000~K}, and $\beta$ from 0 to 2). We repeated the $\chi^2$ minimization, locating the new best-fit parameters for both the star and the hot dust components.
\item As a next step, we refined the fit over the two previous ranges, ensuring consistency between both components and further constraining the parameter space around previously found minima, thus obtaining a new set of best-fitting parameters.
\item The above steps were repeated iteratively incorporating the remaining components at the other wavelength ranges, until the whole SED is fitted and the best-fit parameters for all the components are determined.
\end{itemize}

\subsection{Fitting results}

The NIKA2 upper limits of MN87 are not sufficient to constrain all the SED components keeping a physical meaning. On one hand, the strong infrared emission of this source requires a very large emissivity index ($\beta\gg2$) to obtain the necessary drop at NIKA2 frequencies. On the other hand, the $\approx 10$~mJy level of the radio emission requests a deeply negative value of $\alpha$. In order to find some explanation, both parts of the spectrum must be taken into account; it is known that non-thermal processes are important in this source \citep{ing16}, and the mid-infrared SED points may be heavily contaminated by background emission.

\begin{figure*}
    \centering
    \includegraphics[width=\textwidth]{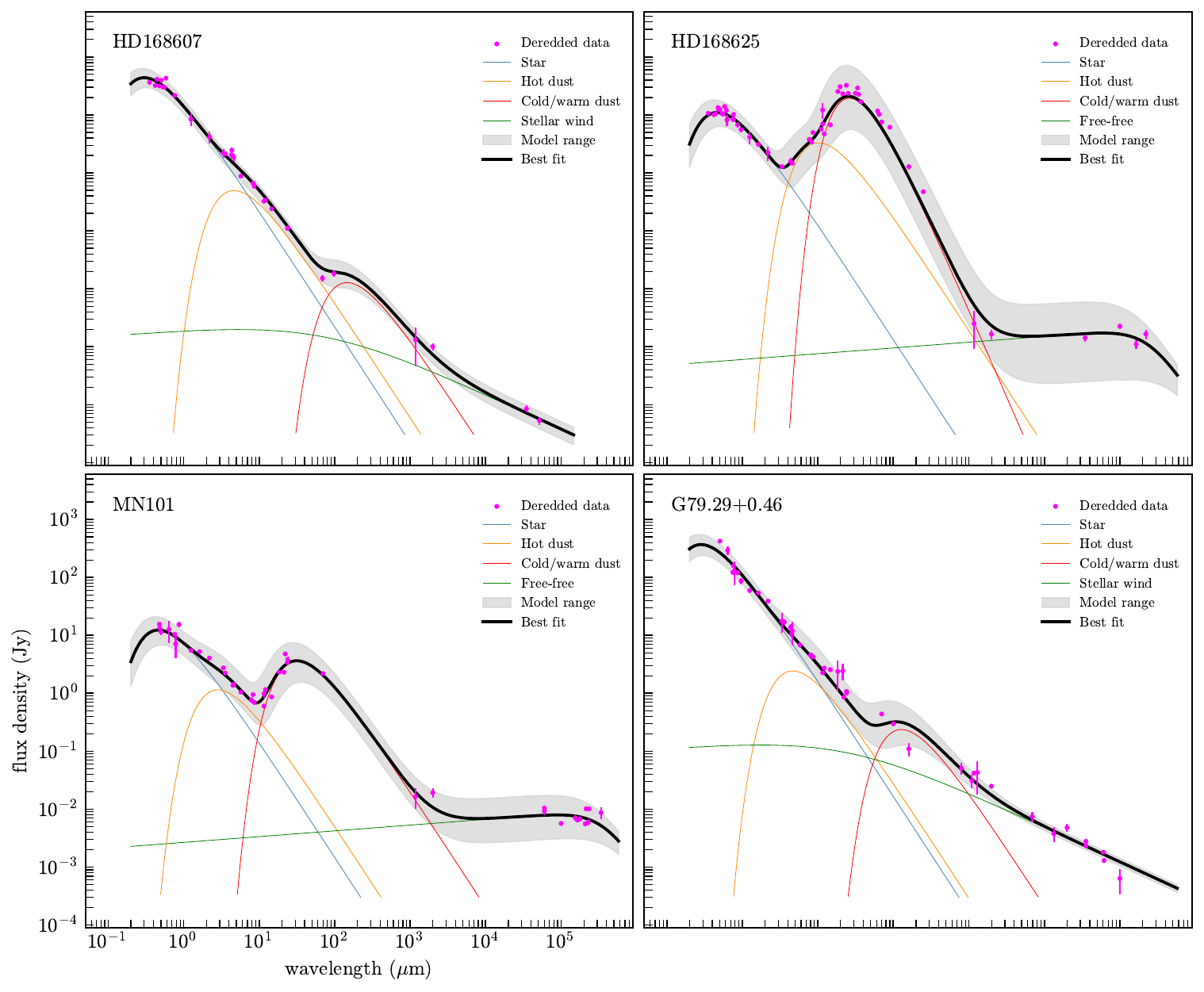}
    \caption{
    Best-fit models. Target names are indicated on the top left corners of each panel. Individual components --star, hot dust, warm/cold dust, and stellar wind / Bremsstrahlung-- are indicated by the blue, orange, red, and green lines, respectively. The combination of the models (best-fit) are shown in bold black lines, being the grey area the one enclosing the combined uncertainties of the fitting. Magenta dots with error bars depict the de-reddened photometry. The four charts are depicted with the same wavelength and flux density scales to ease the comparison among the targets.
    }
    \label{fig:best-fit}
\end{figure*}

For the other stars, Table~\ref{tab:best_fit} summarises the best-fit results. It includes the fundamental stellar parameters ($T_\mathrm{eff}$ and radius $R_\mathrm{star}$), interstellar and circumstellar absorption in $g$ band, parameters of the dust components (dust temperature $T_\mathrm{dust}$, characteristic radius $R_\mathrm{dust}$, and total dust mass $M_\mathrm{dust}$) and, depending on the case, the current mass loss rate $\dot{M}$ of the stellar wind, or the parameters of the free-free model (electron density $n_{\rm e}$ and size of the ionized region $R_{\rm ff}$). Furthermore, Fig.~\ref{fig:best-fit} displays the four best-fit models, together with the individual components and the de-reddened photometry. The four SEDs are plotted at the same scale to ease a visual comparison.

Noteworthy, the two stars with the highest $T_{\rm eff}$, \mbox{HD168607} and G79.29+0.46, are also the two where we fitted their SEDs by including a stellar wind component. Contrarily, \mbox{HD168625} and MN101 were fitted by the low-end of the usual values of $T_{\rm eff}$, and both were fitted by free-free components. The circumstellar extinction is high in all cases, being \mbox{HD168607} the less reddened of the sample. The warm/cold dust masses are remarkably similar, of the order of $10^{-3}\ {\rm M}_\odot$. The hot dust masses, however, are three orders of magnitude lower, except the much larger $M_\mathrm{dust}$ of \mbox{HD168625}, in line with the huge IR emission from its dust nebula. The mass-loss rates inferred from the stellar winds of \mbox{HD168607} and G79.29+0.46 are of the order of a few \mbox{$10^{-6}$~M$_\odot$\,yr$^{-1}$} (scaled to \mbox{$v_\infty=100$ km\,s$^{-1}$}; see Appendix~\ref{app:models} for details).

In the four cases, the relevance of the NIKA2 data is evident. At 1--2~mm, it is clearly noted that the observed flux results as the combined contribution from the ionised stellar wind (or ionised nebula) and the cold dust. The NIKA2 fluxes are vital to constrain the relative contribution of both components arising from the far infrared and from cm-wavelengths.

Below, we compare our fitting results with those available in the literature.

\paragraph{HD168607.} 
The best-fitting model predicts an effective temperature of 17000~K, higher than the value of  $\approx 10^4$\,K determined by \citet{cla12}. The luminosity resulting from our fitting is $10^{5.88\pm0.19}$~L$_\odot$, in good agreement with the estimates provided in the same work, and inferred from a visual inspection to one of its figures. There is a hot dust component very close to the star ($T_\mathrm{dust} = 1100$~K), responsible for most of the emission in the near-infrared, and a more extended --but still unresolved-- cold dust component (35 K) that causes the bump observed around 100$\mu$m. The radio part of the SED is well-fitted by a stellar wind. \cite{lei95}, based on \mbox{8.64 GHz} continuum, determined \mbox{$\dot M=2.3\pm0.5\times10^{-6}$ M$_\odot$ yr$^{-1}$}, using 2.2~kpc and \mbox{140~km\,s$^{-1}$} for the distance and for $v_\infty$, respectively. Their value of $\dot M$ corrected by the distance of 1.84~kpc shifts to \mbox{$1.5\pm0.3\times10^{-6}$ M$_\odot$ yr$^{-1}$}. On the other hand, if we correct our $\dot M$ considering \mbox{$v_\infty=$140~km\,s$^{-1}$}, we obtain 
\mbox{$1.4\pm0.4\times10^{-6}$ M$_\odot$ yr$^{-1}$}, in perfect agreement with \citet{lei95}.

\paragraph{HD168625.} 
According to Table \ref{tab:best_fit}, the luminosity derived from our model is $10^{4.93\pm0.27}\,{\rm L}_\odot$. The strong infrared emission was fitted by a combination of a black and a grey bodies. The black body corresponds to a compact dust component of $T_{\rm dust} = 500$~K (by far below the 1000--2000~K of the other targets), while the grey body represents a more extended warm dust component ($T_{\rm dust} = 140$~K). This is the only source in which a grey body component was required to fit the data, as it is the only way to reconcile the far-infrared emission with the NIKA2 millimetre measurements. 

The best-fit emissivity index, $\beta=1.0\pm0.2$, brings some hints about the dust. Theoretical models \citep{dra84} predict $\beta$ values between 1 and 2, with the larger values in the diffuse ISM \citep{pla14}, and the lower ones in dense cores \citep{for14} and circumstellar disks \citep{fri18}. Physical conditions in the outskirts of evolved massive stars differ markedly from those in the ISM, characterized by diffuse radiation fields, low temperatures, and low densities. The close environs of these hot stars are bathed by strong UV radiation fields but, at the same time, self-shielding is possible due to the continuous mass injection through the dense stellar winds.

Therefore, the value of $\beta$ derived from the fitting is indicative of large grains formed through continuous accretion, favoured by the current photospheric temperature. Indeed, \cite{koc11} proposed that episodes of enhanced mass-loss could be produced during the cold phase of LBV stars and create appropriate conditions (in terms of density and self-shielding) for grain growth. This scenario is further supported by the study of the dust around the LBV HR~Car, performed by \citet{bue17}, which revealed a radial gradient in $\beta$, with lower values near the star and an increase at larger distances, where smaller grains mix with the ISM.

In order to proceed to a fair comparison of our estimates with other published parameters, it must be considered that the assumed distance to \object{HD168625} may significantly differ from one paper to another, with obvious systematic deviations of the quoted values. We therefore re-scaled the published parameters to 1.53~kpc (Table \ref{tab:sources}) when necessary. 

\citet{oha03} imaged the nebula at several near- and mid-IR wavelengths, assuming a distance of 2.8 kpc. After scaling their distance to ours, their fundamental stellar parameters result in close agreement with our estimates: $T_{\rm eff}=14000$~K and $R_{\rm star}=82$~M$_\odot$. They also obtained a dust mass of $\approx 7.5\times10^{-4}$~M$_\odot$, again after correcting to our distance. This is somewhat smaller than the values shown in Table \ref{tab:best_fit}, but still within the uncertainties. Even so, the difference may indicate a real value of $\kappa_0$ a bit higher than the assumed in this work, probably more close to  \mbox{0.9~cm$^2$\,gr$^{-1}$}. 

The stellar luminosity computed by us is substantially lower than that of \citet{mah16}, which is $10^{5.58\pm0.11}\,{\rm L}_\odot$ also assuming a distance of 2.8~kpc. This difference is impossible to be explained in terms of uncertainties. However, after correcting their value to a distance of 1.53 kpc, we get $\sim 10^{5.05}\,{\rm L}_\odot$, a result fully compatible with our luminosity estimates. In the next Section, we discuss the implications of this revised lower luminosity.

Our warm dust temperature is in accordance with several studies. \citet{oha03} found dramatic spatial changes of the physical dust properties in the nebula, with a representative value of \mbox{$T_{\rm dust}=120$\,K}. Important temperature variations were also found by \citet{uma10}, with values between 110 and 210~K (one of the figures impressively show a large dust plateau with a characteristic temperature of $\approx 120$~K). \citet{blo14} included this source in a sample of evolved stars observed with Herschel/PACS at the \mbox{69\,$\mu$m} olivine band, finding a strong signature of the band and determining a dust temperature range of 50 to 150~K. Furthermore, \citet{arn18} made a sophisticated dust model using mainly data from SOFIA and determined \mbox{$T_{\rm dust}=70\pm40$~K} and \mbox{$M_{\rm dust}=2.5\pm 0.1\times 10^{-3}$~M$_\odot$}. Summing up, the values of the cold dust component quoted in this work are in remarkable agreement with the above works and reinforce the validity and reliability of the fitting.

The stellar wind model of \citet{uma10} is highly appropriate to separate the star contribution from the total flux (i.e., star plus nebula) at cm-wavelengths, but it is not able to account for the (relatively low) NIKA2 fluxes. Furthermore, a bright and highly ionized nebula is known to be excited by \mbox{HD168625}. We therefore proceeded to model the ionized gas by a free-free component, which (in combination with the grey body) successfully explained the SED at far-infrared, mm-, and cm-wavelengths.

\paragraph{MN101.}
At the best of our knowledge, there are no estimates of the fundamental stellar parameters of MN101. Our best-fit $T_{\rm eff}$ is exactly equal to that of \mbox{HD168625}, while $R_{\rm star}$ is slightly smaller. These parameters are compatible with a low-luminosity, relatively cold LBV  \citep{smi04}. The strong infrared emission, peaking in the \mbox{$20-30\,\mu$m} range, is well modelled by an au-size, hot dust component at \mbox{1800\,K}, plus a colder one, at \mbox{160\,K}, extending up to 100~au. Considering the similarity between MN101 and \mbox{HD168625}, we modelled the cm-wavelength emission using a simple free-free model, which reconciles the different contributions in the vicinity of the star. In the Fig.~\ref{fig:best-fit}, we note that both the thermal 160~K dust and the free-free emission from the ionised gas provide comparable fluxes to the total millimetre emission observed by NIKA2. 

We emphasize that this fitting constitutes just a first (and rather crude) approximation. In any case, it provides the first estimates of the fundamental star parameters and some global values of its closest circumstellar material (the good coincidence of our parameters in \mbox{HD168625} and those reported in the literature gives support to our approach). Obviously, other different mechanisms may be at play in the star surroundings \citep[e.g, strong stellar winds and non-thermal emission;][]{ing16}, but they are not significant enough compared with the 160~K dust plus the ionised region associated with the nebula, in the cm- and mm-wavelength domains.

\paragraph{G79.29+0.46.} 
The best-fit stellar parameters of this source are similar to those of \mbox{HD168607}. We computed a luminosity of $10^{5.78\pm0.21}$~L$_\odot$. This luminosity and the fitted $T_{\rm eff}$ (18\,000\,K) puts this target in the middle of the S~Doradus instability strip of the HR diagram \citep{smi04}. This luminosity is also close to that quoted by \citet{agl14}, $10^{5.4}$~L$_\odot$, in a study focused to fit the SED in the visual and near infrared. 

\citet{wat96}, based on infrared spectroscopy, determined a mass-loss rate of \mbox{$\approx 10^{-6}$~M$_\odot$\,yr$^{-1}$} (for a measured wind velocity of \mbox{94\,km\,s$^{-1}$}) in excellent agreement with our values. Wind velocity was also measured by \citet{voo00}, who determined a value of \mbox{110\,km\,s$^{-1}$}; therefore, the $\dot{M}$ quoted in Table~\ref{tab:best_fit} seems close to the real one, considering that it is scaled to \mbox{100\,km\,s$^{-1}$} of wind velocity. It is worth noting that, in the work of \citet{voo00}, the necessary stellar input to explain the infrared brightness distribution is also in close agreement with our estimates: $T_{\rm eff}=20\,000$~K and $R_{\rm star}=76~{\rm R}_\odot$ (equivalent to a luminosity of $10^{5.92}$~L$_\odot$).

\citet{kra10} also modelled the infrared part of the central source and determined a dust temperature of 1800~K. This is not far from our results, but these authors considered only data up to \mbox{$24\,\mu$m} and did not attempt to include those at larger wavelengths. Contrarily, \citet{agl14} fitted the whole infrared range by summing up five grey bodies adopting $\beta=2$ for all of them; the temperature of these components ranges from 40 to 1200~K, a result consistent with our two (hot and cold) dust components.

\begin{table*}
\caption[]{Best-fit model results.}
\label{tab:best_fit}
\centering
\begin{tabular}{ccccccccccc}
\hline\hline
\noalign{\smallskip}
Source & $T_\mathrm{eff}$ & $R_\mathrm{star}$ & $A_\mathrm{g}^\mathrm {ism}$ & $A_\mathrm{g}^\mathrm {csm}$ & $T_\mathrm{dust}$ & $R_\mathrm{dust}$ & $M_\mathrm{dust}$ & $\dot{M}$\tablefootmark{(a)} & $n_\mathrm{e}$ & $R_\mathrm{ff}$ \\
\noalign{\smallskip}
& $10^3$\,K & R$_\odot$ & mag & mag & K & au & M$_\odot$ & $10^{-6}$\,M$_\odot$\,yr$^{-1}$ & cm$^{-3}$ & $10^3$\,au \\
\noalign{\smallskip}
\hline
\noalign{\smallskip}
HD168607 & $17\pm1$ & $100\pm10$ & 4.71 & 1.6 & $1100\pm100$ & $3\pm1$             & $6.4\,10^{-6}$ & $1.0\pm0.3$   \\
           &          &            &      &     & $35\pm5$     & $85\pm5$            & $5.1\,10^{-3}$                 \\
\noalign{\smallskip}
HD168625 & $11\pm1$ &  $80\pm10$ & 1.13 & 4.2 & $500\pm100$  & $21\pm1$            & $3.1\,10^{-4}$ & & $2\pm1\,10^4$ & $1.2\pm0.2$ \\
           &          &            &      &     & $140\pm20$   & ---\tablefootmark{(b)} & $2.0\,10^{-3}$  \\
\noalign{\smallskip}
MN101      & $11\pm1$ & $70\pm10$  & 7.56 & 8.3 & $1800\pm200$ & $1.5\pm0.5$         & $1.6\,10^{-6}$ & & $8\pm2\,10^3$ & $3.2\pm0.4$ \\
           &          &            &      &     & $160\pm20$   & $100\pm20$          & $7.1\,10^{-3}$ \\
\noalign{\smallskip}
G79.29+0.46& $18\pm1$ & $80\pm10$  & 9.96 & 5.2 & $1100\pm100$ & $2\pm1$             & $2.8\,10^{-6}$ & $2.2\pm0.3$ \\
           &          &            &      &     &    $40\pm10$ & $90\pm10$           & $5.7\,10^{-3}$ \\
\noalign{\smallskip}
\hline
\end{tabular}
\tablefoot{
\tablefoottext{a}{
A wind velocity of 100 km\,s$^{-1}$ is assumed (Appendix~\ref{app:models}.
}
\tablefoottext{b}{Not possible to estimate due to opacity (see Sect.~\ref{sec:fitting} and Appendix~\ref{app:models}).
}}
\end{table*}

\section{Concluding remarks}

We detected the central sources in four out of the five surveyed LBVs at both mm bands. We also found extended emission which reaches up to some 1--2~pc from the stars, although not all the features seems evidently associated with the targets. This is a step forward in the field because the mm-continuum has been observed only in a handful of massive stars \citep{lei91, hig94, agl17b, agl19, fen18}.  

With the 30m beam size at the NIKA2 frequencies, our data can resolve out structures typically larger than \mbox{$\sim10^4$\,}au at the distances of our targets. We distinguished in the mm maps several features: (1) unresolved sources, corresponding to the stellar winds, probably enshrouded by ionised gas and dust; (2) a plateau of extended emission, particularly noted in HD~168625; (3) a totally detached shell in G79.29+0.46; (4) some Galactic clouds, not clearly linked to the stars; (5) the IRDC in the G79.29+0.46 field.

These features may be characterized not only by their morphology, but also by their spectral indices ($\alpha_{\rm mm}$). For the point sources, except MN101, $\alpha_{\rm mm}$ are close to 0.6, which is the expected result for a ``canonical'' stellar wind. This trend becomes different when computed $\alpha_{\rm mm}$ for the extended emission. The Galactic clouds in the fields of MN87 and MN101 and the IRDC in the field of G79.29+0.46 have the highest values of $\alpha_{\rm mm}$ (well above 2.5), typical of thermal dust emission. On the opposite end, there are the negative values found in the close environs of the central sources, where the stellar wind contribution seems diluted by more extended nebular material. In the field of G79.29+0.46, we noted two areas with clear gradients of $\alpha_{\rm mm}$, from negative to positive values when moving outwards the star position; we propose an explanation of this behaviour by some kind of low-velocity shocks acting in one of the well-known shells \citep{riz08}, and also in a region of interaction with the IRDC \citep{pal14}.

In order to provide a better understanding of the relative importance of the excitation mechanisms, we complemented our NIKA2 data by an exhaustive search for archives, surveys, and dedicated observations available in the Virtual Observatory. For each target, we constructed six-decade SEDs with information from the visual to the radio regimes.

After a careful curation of the data, we distinguished two kinds of SEDs, which are determined by the relative flux density at mm-wavelengths compared to their cm counterparts. The main feature of the first group (to which HD~168607 and G79.29+0.46 belong) is that the mm flux density is significantly higher (one order of magnitude) than the cm flux. In turn, the second group is characterized by comparable flux densities at both mm- and cm-wavelengths. In the five cases observed, the SEDs show that the sources are highly reddened and depict strong near- and mid-infrared emission.

We also proceeded to do a (somewhat generic) modelling of the SEDs, with the objective of identifying different physical components and their relative contribution. The approach used kept the number of components and parameters as small as possible. In this way, we accomplished to confirm or predict the existence of the modelled components, without including sophisticated information about structure, composition, or any other properties.

The fitting is overall satisfactory, considering that the data are spanning six decades of the electromagnetic spectrum, the diversity of the instruments included, and the minimal number of free parameters employed. Notably, it allowed us to unveil the presence of unresolved hot dust components at only 4 to 6 times the stellar radius in all cases but \mbox{HD168625}, where the 500~K gas may extend up to 50--60 $R_{\rm star}$. It is known that such close-in structures are found in other massive evolved stars, like B[e] supergiants, surrounded by circumstellar disks \cite{zick85}. In the context of LBV stars, the detection of au-sized hot dust structures (500 to 2000~K) around both active and quiescent sources may be explained by a dynamics where the ejecta expelled during a past mass-loss event falls back to the star, as proposed by some hydrodynamical simulations \citep{owo13, owo19}. Such fallback would produce dense rotating disks where dust grains can survive against the hot radiation from the central star and grow efficiently, helped by self-shielding. Of course, very high-resolution IR and mm observations are needed to test this plausible but highly speculative interpretation.

In \object{HD168607} the mass loss rate was better constrained thanks to the combination of a better knowledge of its distance, the use of a measured wind velocity, and a fitting which considered the contribution of more extended cold dust. 

The revised luminosity of \object{HD168625} would place it in the lowest part of the LBV zone in the HR diagram \citep{smi04}. This star has long been regarded as a ``twin'' of \mbox{Sk\,-69\,202}, the blue supergiant identified as the progenitor of SN1987A \citep{wal87}, owing to the striking similarities of their circumstellar material \citep{smi07}. With the revised parameters made in this work, the resemblance of HD168625 and \mbox{Sk\,-69\,202} is even stronger: HD168625 is nearly at the same position in the HR diagram as it was \mbox{Sk\,-69\,202} before the SN explosion \citep[][see Fig.~6]{sma02}, i.e., with nearly identical physical properties. 

The non-detection of MN87 is intriguing. The upper limits for the central source at both NIKA2 frequencies are well below the expected fluxes from thermal cold dust or ionized gas. While a non-thermal contribution could help explain such low fluxes, its inclusion is challenging considering the significant IR emission and the thermal spectral index of the central source at cm-wavelengths \citep{ing16}. 

This work presents the first estimates of the fundamental parameters of the central source of MN101. These results are compatible with a LBV close to its coldest value \mbox{($T_{\rm eff}=11\,000$~K)}. Its SED is very similar to that of HD168625, although its hot dust component is the most compact \mbox{($1-2$~au)} and hottest of the sample \mbox{($T_{\rm dust}=1800$~K)}.

In G79.29+0.46, the submm/mm/cm-wavelength flux (and consequently the energy input from the hot star) remained stationary during at least 22 years \citep[see][and our Appendix~\ref{app:notes}]{hig94, cro07, uma10}. In the absence of a systematic photometric monitoring of G79.29+0.46, the apparent steadiness of the stellar wind and the dust properties brings strong support to the ``candidate'' LBV status of this source \citep{voo00}. 

It is remarkable that the stars which share a given ``type'' of SED also display similar $T_{\rm eff}$. Specifically, the two stars with a monotonically decrease from infrared to cm ranges (HD168607 and G79.29+0.46) have the highest $T_{\rm eff}$ and their mm and cm SEDs are well-fitted only by stellar winds, while the other sources depict the opposite behaviour (low $T_{\rm eff}$ and free-free components). There is not a clear explanation of this effect, although the sample is admittedly small to bring any conclusion.

This pilot study demonstrates the importance of studying the mm-wavelength continuum emission in this kind of objects. The NIKA2 frequencies fall between the far-infrared and the radio domains, which are overall constrained by cold dust thermal emission and Bremsstrahlung mechanism, respectively. Therefore, NIKA2 frequencies are very well suited to understand in detail the relative contribution of these two mechanisms, with the consequent determination of important parameters about the energy injected by the LBVs to their surrounding circumstellar material. Likewise, it suggests the existence of different "families" of SEDs, closely tied to the stellar parameters. Follow-up works addressing a larger, statistically significant sample of LBV stars can shed further light on these aspects.

\begin{acknowledgements}
We would like to thank the IRAM staff for their kind and professional support during the observations and the NIKA2 core team for providing the data analysis software that was used to reconstruct the maps shown in the article. J.R.R. acknowledges support by PID2022-137779OB-C41 funded by MCIN/AEI/10.13039/501100011033 by ``ERDF A way of making Europe''. This research has made use of the Spanish Virtual Observatory (\mbox{\url{https://svo.cab.inta-csic.es}}) project funded by MCIU/AEI/10.13039/501100011033/ through grant PID2023-146210NB-I00.
\end{acknowledgements}

\bibliographystyle{aa}
\bibliography{references}

\begin{appendix}

\section{Spectral energy distributions}
\subsection{Data origin}
\label{app:surveys}

In Table \ref{tab:catalogues} we detail basic information about the catalogues, surveys, and observations used to build the SEDs.

\begin{table}[h!]
\caption[]{Surveys and catalogues.}
\label{tab:catalogues}
\centering
\begin{tabular}{llcc}
\hline\hline
\noalign{\smallskip}
Name & VizieR ID & Range & Ref. \\
\noalign{\smallskip}
\hline
\noalign{\smallskip}
Hipparcos     & I/239           & UV-opt      &  1 \\
TIC82         & IV/39           & UV-opt      &  2 \\
SDSS          & I/305; I/353    & optical     &  3 \\
PAN-STARRS    & II/349          & opt-IR      &  4 \\
Gaia DR3      & I/355           & optical     &  5 \\
2MASS         & II/246          & near IR     &  6 \\
WISE          & II/311          & near-mid IR &  7 \\
allWISE       & II/328          & mid IR      &  8 \\
unWISE        & II/363          & near IR     &  9 \\
catWISE       & II/365          & near IR     & 10 \\
Spitzer/IRAC  & II/368          & near IR     & 11 \\
GLIMPSE       & II/293          & near IR     & 12 \\
MSX           & V/114           & near-mid IR & 13 \\
SOFIA         & \multicolumn{1}{l}{---} & mid IR & 14 \\
AKARI         & II/297          & near-mid IR & 15 \\
IRAS          & II/125          & mid IR      & 16 \\
MIPSGAL       & J/AJ/149/64     & mid IR      & 17 \\
Hi-GAL        & J/A+A/591/A149  & mid-far IR  & 18 \\
              & J/MNRAS/471/100 &             & 19 \\
Miz10         & J/AJ/139/1542   & mid-IR      & 20 \\
Herschel/PACS & VIII/106/       & far IR      & 21\tablefootmark{a} \\
Hig94         & \multicolumn{1}{l}{---} & submm-mm & 22 \\
{\bf NIKA2}   & \multicolumn{1}{l}{---} & mm    &    \\
NVAS          & \multicolumn{1}{l}{---} & mm-cm & 23\tablefootmark{b} \\
Lei95         & \multicolumn{1}{l}{---} & cm  & 24 \\
Agl14         & \multicolumn{1}{l}{---} & cm  & 25 \\
CORNISH       & J/ApJS/205/1    & cm & 26 \\
Ing14         & \multicolumn{1}{l}{---} & cm  & 27 \\
VLASS         & J/ApJS/255/30   & cm & 28 \\
THOR          & \multicolumn{1}{l}{---} & cm  & 29 \\
MAGPIS        & \multicolumn{1}{l}{---} & cm  & 30 \\
SMGPS         & J/A+A/695/A144  & cm  & 31 \\
RACS          & J/other/PASA/38.58 & cm & 32 \\
GLOSTAR       & J/A+A/680/A92   & cm & 33 \\
\hline
\end{tabular}
\tablefoot{
All the NIKA2 flux densities are derived in this work.\\
\tablefoottext{a}{Flux densities computed in this work from downloaded images at 
\url{http://archives.esac.esa.int/hsa/whsa/}}\\
\tablefoottext{b}{Flux densities computed in this work from downloaded images at 
\url{http://www.vla.nrao.edu/astro/nvas/}}
}
\tablebib{
(1)~\citet{esa97}; (2)~\citet{pae22}; (3)~\citet{las08}; (4)~\citet{cha18}; 
(5)~\citet{gai22}; (6)~\citet{skr06}; (7)~\citet{cut12}; (8)~\citet{cut13}; 
(9)~\citet{sch19}; (10)~\citet{mar21}; (11)~\citet{spi21}; (12)~\citet{spi09}; 
(13)~\citet{ega03}; (14)~\citet{arn18}; (15)~\citet{ish10}; (16)~\citet{hel88}; 
(17)~\citet{gut15}; (18)~\citet{mol16}; (19)~\citet{eli17}; (20)~\citet{miz10}; 
(21)~\citet{mar21}; (22)~\citet{hig94}; (23)~\citet{cro07}; (24)~\citet{lei95}; 
(25)~\citet{agl14}; (26)~\citet{pur13}; (27)~\citet{ing14}; (28)~\citet{gor21}; 
(29)~\citet{wan20}; (30)~\citet{hel06}; (31)~\citet{bor25}; (32)~\citet{hal21}; 
(33)~\citet{yan23}
}
\end{table}

\subsection{Data notes on individual sources}
\label{app:notes}

In this section we report some particular notes which complement the information presented in the SEDs.

\subsection*{HD168607 (Figure \ref{fig:sed_HD168607})}
\begin{figure}[h!]
\centering
\includegraphics[width=1\linewidth]{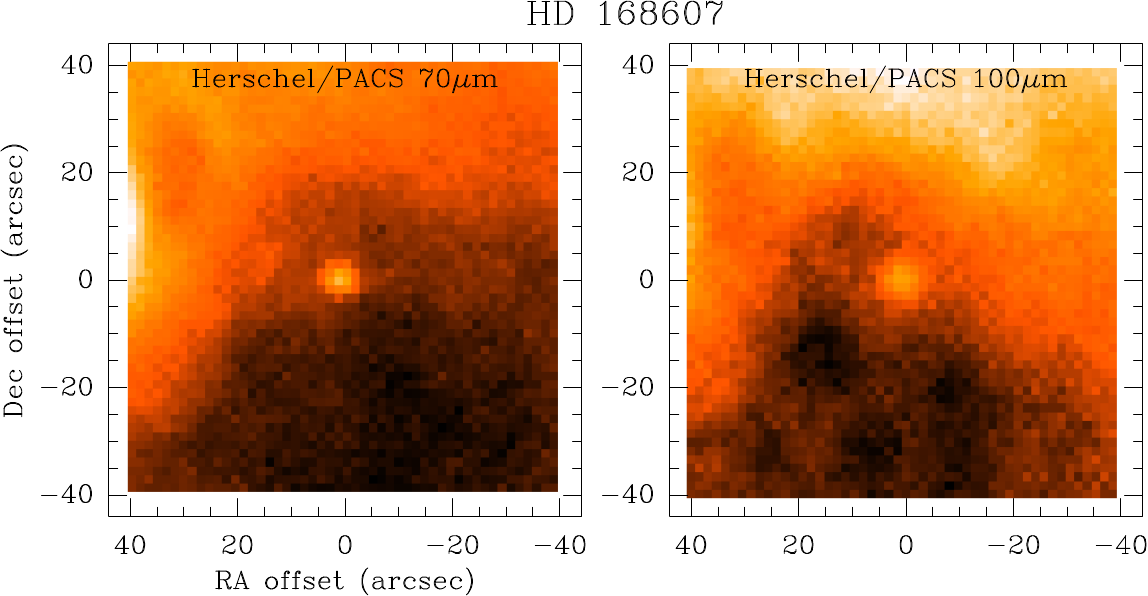}
\caption{Herschel/PACS images around HD168607 at 70\,$\mu$m (left) and 100\,$\mu$m (right). Equatorial coordinates are relative to the star position. The continuum emission at these wavelengths is reported for the first time in this work.}
\label{fig:HD168607_PACS}
\end{figure}

\begin{itemize}
\item Pan-STARRS photometry shows significant dispersion (up to a factor of 4 in $g$) so it was discarded.
\item The WISE band 4 measurement is heavily contaminated by nearby extended emission not related to the star. As no reliable flux could be determined, the measurement was discarded.
\item We found point like emission at 70 and 100\,$\mu$m with Herschel/PACS. The nearby HD168625 is a target of a project (observation IDs from 1342217763 to 1342217766), and the observed images include by chance the position of HD168607. We extracted a field around HD168607 from the original images and fitted a two-dimensional Gaussian source. The results are shown in Fig.~\ref{fig:HD168607_PACS} and in Table \ref{tab:HD168607_fitting}. At \mbox{160\,$\mu$m} the star is no longer detected.
\item This source is catalogued in the GLOSTAR survey at \mbox{1.4\,GHz} \citep{yan23}, but not in CORNISH, MAGPIS, or THOR. In the search for references not available in VizieR, we found a survey made with ATCA at 8.54\,GHz \citep{lei95}, which was added to the SED.
\end{itemize}

\begin{table}[ht]
\caption[]{Fitting of the PACS mid-IR fluxes of \object{HD168607}.}
\label{tab:HD168607_fitting}
\centering
\begin{tabular}{lr}
\hline\hline
\noalign{\smallskip}
Parameter & \multicolumn{1}{c}{Value}\\
\noalign{\smallskip}
\hline
\noalign{\smallskip}
Centroid: \\
Right Ascension  & $275.31\pm0.02$\,deg \\
Declination      & $-16.375\pm0.007$\,deg 
\vspace{2.5mm} \\
70\,$\mu$m: \\
$r_\mathrm{maj}$ & $3.3\pm0.5$\,arcsec\\
$r_\mathrm{min}$ & $2.6\pm0.4$\,arcsec\\
Flux density & $152\pm19$\,mJy
\vspace{2.5mm} \\
100\,$\mu$m: \\
$r_\mathrm{maj}$ & $5.0\pm0.6$\,arcsec\\
$r_\mathrm{min}$ & $4.6\pm0.6$\,arcsec\\
Flux density & $183\pm24$\,mJy\\
\hline
\end{tabular}
\end{table}

\subsection*{HD168625 (Figure \ref{fig:sed_HD168625})}
\begin{itemize}
    \item The source is included in the IRAS Point Source Catalogue, although not detected at \mbox{100\,$\mu$m}. 
    \item In the PACS images at 70 and \mbox{100\,$\mu$m} the star is not detected. In Hi-Gal, the nebula is listed as two different sources (the brightest parts of the nebula); we consequently summed up the two fluxes, and assumed uncertainties of 10\%.
    \item A complete study in the infrared was done by \citet{arn18}, who included their SOFIA measurements together with 2MASS, AKARI, WISE, MSX, Spitzer and Herschel data. We cross-checked the spectral fluxes in the paper with our ones obtained from VizieR, and added the SOFIA and Herschel/PACS data to our SED.
    \item The WISE band 4 flux is clearly overestimated, up to three times the fluxes of MIPS, MSX, IRAS and AKARI in nearby wavelengths. Such effect is probably due to the large aperture of WISE at $22\,\mu$m, which is critical in presence of the strong IR nebula. We therefore remove this point from the SED.
    \item The source is not included in the THOR catalogue, but we downloaded the images and detected some extended emission. We averaged three frequencies (1690, 1820 and \mbox{1950\,MHz}) to get a reliable source detection, and incorporated the average to the SED. This source was also observed by \citet{lei95} at \mbox{8.54\,GHz} using ATCA, and we therefore also included those measured values.
    \item \citet{uma10} observed this target at high resolution with the VLA at 8.54 GHz, detecting the central point source (\mbox{0.42\,mJy}) while resolving out most of the extended nebula; we therefore decided to take this measurement into account for the fitting, as explained below. The VLASS survey also detected this point source at 2--4 GHz, with a surrounding ring of 5--10 arcsec.
    \item After a close inspection to the data, we noted that the THOR flux is higher than the adjacent ones. We explored whether this is due to an incorrect flux measurement\footnote{See Lacy et al. 2019, VLASS Project Memo \#13: Pilot and Epoch 1 Quick Look Data Release \url{https://library.nrao.edu/public/memos/vla/vlass/VLASS_013.pdf.}}, or due to the possibility of reaching the turnover frequency close to the VLASS frequency (3 GHz). To do so, we downloaded and analysed the individual images available in the THOR Image Server\footnote{\url{https://thorserver.mpia.de/thor/image-server/}}. These images correspond to epochs 1.2 and 2.2, but only in the first one the source is detected. Our measured flux is compatible with that provided in the archive, so we confirm its validity and used the tabulated value and its uncertainty.
\end{itemize}

\subsection*{MN101 (Figure \ref{fig:sed_MN101})}
\begin{itemize}
    \item The Gaia G flux is anomalously high with respect to the other Gaia bands, and significantly above measurements of other surveys. This effect may be due to the presence of circumstellar dust which contaminates the star photometry, as explained in Sect.\,\ref{sec:clarification}.
    \item This source is included in most of the radio surveys searched. The flux densities from THOR were quite discrepant with respect to others, so we recomputed them using aperture photometry. Our values are more consistent with the other surveys, and we therefore decided to incorporate them to the SED, instead of the catalogued ones.
    \item The nebular feature was observed with the VLA \citep{ing14} at 1.4 and 5 GHz, with flux densities of $10.2\pm0.8$ and $10.5\pm0.1$~mJy, respectively. Considering that the angular extent of the source ($\sim$20~arcsec) is comparable to the NIKA2 beam, we preferred these values for the fitting. Higher resolution 5~GHz observations by \citet{ing16} revealed a central point source ($7.0\pm0.2$~mJy), with a spectral index compatible with a thermal stellar wind and an ionised circumstellar material characterised by a non-thermal spectral index.
\end{itemize}

\subsection*{G79.29+0.46 (Figure \ref{fig:sed_G79})}
\begin{itemize}
\item The source is highly reddened. G79.29+0.46 has been observed in three different epochs by SDSS; the measurements differ up to a factor of two, probably due to stellar variability. The PAN-STARRS survey, made some years after the SDSS data release, shows a similar behaviour. In fact, G79.29+0.46 is included in the catalogues of \citet{hei18} and \citet{che20} of variable stars. We decided to keep different points for both SDSS and PAN-STARRS to reflect the source variability.
\item \citet[][referred to as Agl14 in the SED]{agl14} observed this source at 8.46, 4.96, and 1.4 GHz using the VLA interferometer, and made a SED study in the infrared domain using observations of Herschel, Spitzer, ISO and WISE. We cross-checked our values in common (e.g., from WISE and 2MASS) and completed our SED by incorporating their values from Spitzer and Herschel. 
\item VLASS does not include this source in its catalogue, although it is noted some point-like emission in the images corresponding to the three epochs observed. We therefore computed the integrated fluxes and averaged them with a inverse-square weighting of the flux uncertainties.
\item The NVAS archive also provides images in the C-, X-, \mbox{K-,} and Q-bands. We computed the integrated fluxes in these images as well, and incorporated them to the SED. The flux uncertainties in C- and X-bands are notoriously large in comparison to those of the targetted observations of \citet{agl14} at similar frequencies; we therefore decided to keep these NVAS data in the SED, but remove them as input for the fitting.
\item \citet{hig94} made a multi-wavelength study of the central source and the surrounding shells, including data from IRAS and targetted observations of several instruments from 0.8~mm to 5~GHz. The fluxes measured at all these frequencies are consistent with our NIKA2 values and the above mentioned radio surveys (NVAS, VLASS, Agl14), which demonstrates that G79.29+0.46 remained stationary at cm- and mm-wavelengths at least during the last 20 years.
\end{itemize}

\section{Modelling details}
\label{app:models}

\subsection*{Central star} 

The contribution of the star was modelled as a black body, parametrised by the effective temperature $T_\mathrm{eff}$, the stellar radius $R_{\rm star}$, and the distance $d$ to the star. In this way, the flux density is:

\begin{equation}
    S_\nu = \Omega \ B_{\nu}(T_{\rm eff}) = \Omega \ \frac{2\ h\ \nu^{3} / c^{2}}{\exp(h\ \nu / k\ T_\mathrm{eff}) - 1}
    \label{eq:bb}
\end{equation}

\noindent with $\Omega$ being the solid angle subtended by the source, \mbox{$\Omega = \pi\ (R_{\rm star}/d)^2$}, and $B_\nu(T_{\rm eff})$ the Planck function.

The simplicity of this model is adequate for the purpose of this article. The uncertainties introduced by this simplification is by far below those due to the variability of the stars and the heterogeneity of the collected visual and near-infrared data. This fitting is sufficient to provide good approximations of $T_\mathrm{eff}$ and $R_{\rm star}$, and to constrain the neighbour range of wavelengths, where the hot dust is dominant.

\subsection*{Circumstellar dust}

The emission arising from this component was modelled by modified black bodies (sometimes referred to as grey bodies), parametrized by a dust temperature $T_{\rm dust}$, a characteristic radius of the emitting region $R_{\rm dust}$ (expected much larger than $R_{\rm star}$), and a dust opacity law \citep{oss94}:

\begin{equation}
\kappa_\nu = \kappa_{0}\ (\nu/\nu_0)^\beta
\end{equation}

In the above equation, $\kappa_\nu$  is the dust mass opacity in units of cm$^2$ g$^{-1}$, $\kappa_{0}$ is the dust opacity at the reference frequency $\nu_0$ (here chosen at 230~GHz) and $\beta$ is the dust emissivity index. The measured flux density of the modified black body results:

\begin{equation}
S_\nu = \sigma\ \kappa_\nu\ {\rm B}_\nu(T_{\rm dust})\ \Omega_{\rm cloud}
\end{equation}

\noindent where $\sigma$ is the dust mass column density along the line of sight (in units of g~cm$^{-2}$) and $\Omega_{\rm cloud}$ the solid angle subtended by the cloud of radius $R_{\rm dust}$. If the total dust mass of the cloud is $M_{\rm dust}$, we can roughly compute

$$
\sigma = \frac{M_{\rm dust}}{\pi\ R_{\rm dust}^2}
$$

$$
\kappa_\nu = \kappa_{\nu_0} (\nu/\nu_0)^\beta = \kappa_{\nu_0} (\lambda_0/\lambda)^\beta
$$

$$
\Omega = \pi\ (R_{\rm dust}/d)^2
$$

\noindent and the flux density results:

\begin{equation}
S_\nu = M_{\rm dust}\ \kappa_0\ \lambda_0^\beta\ \lambda^{-\beta}\ {\rm B}_\nu(T_{\rm dust})\ / d^2
\label{eq:greybody}
\end{equation}

For low to moderate gas densities, dust opacities of MRN grain \citep{mat77} with and without ice mantles ranges from 0.3 to 0.9\,cm$^2$~g$^{-1}$. So far, we estimated the dust masses in Table \ref{tab:best_fit} by assuming $\kappa_0=0.5 \,{\rm cm}^2\,{\rm g}^{-1}$, which yields uncertainties due to unknown $\kappa_0$ within a factor of two. In this way, the can fit the density flux arising from a dust cloud in terms of $\beta$, $M_{\rm dust}$, and $T_{\rm dust}$.

The parameter $\beta$ encapsulates key dust properties, such as composition, porosity, and grain size distribution. Therefore, single-$\beta$ grey bodies are just a simplification without a specific physical meaning \citep{jon13}. Even so, they provide useful information about the dust behaviour. In our modelling we considered $\beta$ as a free parameter, although it was indeed necessary only for the cold component of \mbox{HD168625}, where the best-fit was found for $\beta=1\pm0.2$. For the other cases, the best-fit value of $\beta$ was zero, i.e., all other components are well-fitted as ideal black bodies.

When the best-fit corresponds to a black body, we can additionally derive $R_{\rm dust}$ because the density flux is also:

\begin{equation}
S_\nu = {\rm B}_\nu(T_{\rm dust})\ \Omega = {\rm B}_\nu(T_{\rm dust})\ \pi\ (R_{\rm dust}/d)^2
\end{equation}

In practice, we fitted the SED through the expression:
\begin{equation}
S_\nu = \frac{1}{(\lambda/1\mu\rm m)^\beta}\ {\rm B}_\nu(T_{\rm dust})\ \pi\ (\hat{R}/d)^2
\label{eq:fiducial}
\end{equation}
\noindent where $\hat{R}$ is considered as a fiducial radius. 
By equating Eqs.~\ref{eq:greybody} and \ref{eq:fiducial} we can derive, after some algebra, a relationship between $M_{\rm dust}$ and $\hat{R}$:
\begin{equation}
M_{\rm dust} = \frac{\pi\ \hat{R}^2}{\kappa_0\ (\lambda_0/1\mu\rm m)^\beta}
\end{equation}
When $\beta\ne0$, $\hat{R}$ is just a proxy to derive $M_{\rm dust}$, without a real physical meaning. 
On the other hand, when the best-fit corresponds to a black body ($\beta=0$) $\hat{R}=R_{\rm dust}$. Due to the uncertainties and assumptions involved, the masses must be considered just as a rough approximation.

\subsection*{Stellar wind}

This component was modelled by a stationary, isotropic, and isothermal stellar wind, expanding at constant velocity \citep{pan75,wri75}. In the optically thick regime, the flux density of such a wind at the frequency $\nu$ is:

\begin{equation}
\begin{split}
S_{\nu}\ ({\rm thick}) &= 5.12\, \left(\frac{\nu}{10 \, \text{GHz}}\right)^{0.6} \left(\frac{T_\text{e}}{10^4 \, \text{K}}\right)^{0.1} \left(\frac{\dot{M}}{10^{-5} \, \mathrm{M}_{\odot}/\text{yr}}\right)^{4/3} \\
& \times \left(\frac{\mu}{1.2}\right)^{-4/3} \left(\frac{v_{\infty}}{10^3 \, \text{km/s}}\right)^{-4/3} \left(\bar{Z}\right)^{2/3} \left(\frac{d}{\text{kpc}}\right)^{-2} \ {\rm mJy}
\end{split}
\label{eq:sw}
\end{equation}

\noindent where $T_\text{e}$ is the electronic temperature, 
$\dot{M}$ the mass-loss rate, 
$\mu$ the mean molecular weight (assumed 1.2), 
$v_\infty$ the expansion velocity of the wind, and
$\bar{Z}$ the average ionic charge (assumed 1). 
The optically thick assumption holds up to a certain turnover frequency $\nu_\text{c}$, equal to

\begin{equation}
\nu_c = 4.7 \times 10^{42} \, \frac{\gamma^{1/2}\,g_{\rm ff}^{1/2}\,\bar{Z}}{T_\text{e}^{3/4}\,R_\text{c}^{3/2}} \, \frac{\dot{M}}{\mu \, v_{\infty}} \, \text{Hz}
\label{eq:turnover}
\end{equation}

\noindent with $\gamma$ being the number of electrons per ion and $g_{\rm ff}$ the free-free Gaunt factor, approximated as \citet{lei91}:

\begin{equation}
g_{\rm ff}=9.77\,\left(1 + 0.13\,\log\frac{ 
T_\text{e}^{3/2}}{\bar{Z}\,\nu}\right)
\label{eq:gff}
\end{equation}

\noindent and $R_\text{c}$ denoting the inner boundary radius of the ionised envelope (here assumed equal to $R_{\rm star}$). Above $\nu_\text{c}$ the spectrum begins to flatten, approaching that of an optically thin H\textsc{ii} region with $\alpha=-0.1$. In this regime, the flux density results:

\begin{equation}
S_\nu\ ({\rm thin}) = S_{\nu_c}\ (\nu/{\nu_c})^{-0.1}
\end{equation}

We may note in Eq.~\ref{eq:sw} the weak dependence of the flux density with $T_{\rm e}$. In any case, we tested different models by varying $T_{\rm e}$ from $10^3$ to $10^5$~K and noted that the changes in the flux were always smaller than the uncertainties of the observed fluxes. We therefore adopted $T_{\rm e}=10^4$~K to avoid an unnecessary free parameter. Finally, there is a degeneracy in $\dot M$ and $v_\infty$ because the flux density varies with $(\dot M / v_\infty)^{4/3}$. For simplicity, we adopted \mbox{$v_\infty = 100$ km\,s$^{-1}$} and leave $\dot M$ as the only free parameter.

There is a circular dependence between $\nu_c$ and $g_{\rm ff}$ because $\nu_c$ is frequency-dependent through $g_{\rm ff}$. It is therefore necessary to adopt a fixed value of $g_{\rm ff}$. The Gaunt factor $g_{\rm ff}$ varies roughly from 2 to 6 within its validity range (from cm to mm wavelengths), while $\nu_c$ falls in the mid-IR range. Consequently, we adopted a characteristic $g_{\rm ff} = 4.08$, corresponding to a frequency of 30~GHz, for the calculation of $\nu_c$. This causes an uncertainty in the turnover frequency up to a factor of 2, which has no significant impact on the millimetre/centimetric fluxes.

\subsection*{Bremsstrahlung}

For this component, we modelled the free-free emission from an ionised, static, and isothermal H\textsc{ii} region characterised by a radius $R_{\rm ff}$. In the optically thin regime \citep{mez67}, the flux density behaves as:

\begin{equation}
S_{\nu}\ {\rm (thin)} = 2\ k\ T_\text{e} \left(\frac{\nu}{c}\right)^2 \tau \ \Omega_{\rm ff} 
\label{eq:ff}    
\end{equation}

$\Omega_{\rm ff}=\pi\,(R_{\rm ff}/d)^2$ is the solid angle of the ionised region and $\tau$ is the optical depth, given by

\begin{equation}
\tau = 3.28 \times 10^{-7} \left(\frac{T_\text{e}}{10^4\, \text{K}}\right)^{-1.35} \left(\frac{\nu}{\text{GHz}}\right)^{-2.1} \left(\frac{EM}{\text{pc cm}^{-6}}\right)
\label{eq:ff-tau}
\end{equation}

\noindent where $EM$ represents the emission measure of the ionized region. 
    
Solving Eq. \ref{eq:ff-tau} for $\tau=1$, it is possible to compute the turnover frequency $\nu_\text{c}$. At frequencies below $\nu_\text{c}$, the {\sc Hii} region enters in the optically thick regime, the spectrum steepens, and it follows the Rayleigh-Jeans approximation ($S_\nu\propto\nu^2$).

\end{appendix}
\end{document}